\RequirePackage[log]{snapshot}
\documentclass{egpubl}
\JournalSubmission    
\usepackage[T1]{fontenc}
\usepackage{dfadobe}

\usepackage{cite}  
\BibtexOrBiblatex
\electronicVersion
\PrintedOrElectronic
\ifpdf \usepackage[pdftex]{graphicx} \pdfcompresslevel=9
\else \usepackage[dvips]{graphicx} \fi

\usepackage{egweblnk}

\usepackage{hyperref}

\usepackage{amsmath,amsfonts,amssymb,amscd,amsthm,xspace}
\usepackage{graphicx}
\newcommand{\fref}[1]{Figure~\ref{#1}}

\newcommand{\cref}[1]{Chapter~\ref{#1}}

\newcommand{\sref}[1]{Section~\ref{#1}}

\theoremstyle{plain}

\usepackage{mdframed}
\theoremstyle{definition}
\newtheoremstyle{break}
  {1em}{1em}%
  {}{}%
  {\bfseries}{}%
  {\newline}{}%
\theoremstyle{break}
\newmdtheoremenv{definition}[theorem]{Definition}
\newmdtheoremenv{lsystem}{L-system}
\theoremstyle{remark}

\usepackage[dvipsnames, xcdraw]{xcolor}
\usepackage{soul}

\newcommand{\etal}{et al.\ }

\usepackage{tikz}
\usepackage{tikz-3dplot}
\usetikzlibrary{quotes,angles}
\usetikzlibrary{calc}
\usetikzlibrary{shapes}

\usepackage{pgfplots}
\pgfplotsset{width=10cm,compat=1.9}
\usepackage{pgfplotstable}
\usepackage{esvect}
\usepackage{bm}
\usepackage{xfrac}
\usepackage{xfrac}
\usepackage{gensymb}
\usepackage[super]{nth}
\usepackage{placeins}
\usepackage{subcaption}
\usepackage{mathtools}
\usepackage{algorithmic}
\usepackage[ruled]{algorithm2e}

\DeclareMathOperator{\atantwo}{arctan2}
\newcommand{\norm}[1]{\lvert \lvert #1 \rvert \rvert}
\usepackage{tabularx}
\newcolumntype{L}{>{$}l<{$}}
\newcolumntype{P}[1]{>{\centering\arraybackslash}m{#1}}
\newcolumntype{K}[1]{>{\raggedright\arraybackslash}m{#1}}
\newcolumntype{Q}[1]{>{\raggedleft\arraybackslash}m{#1}}
\usepackage{multirow}
\usepackage{xurl}

\title[Procedural Generation and Rendering of Forests]%
      {Procedural Generation and Rendering of Realistic, Navigable\\ Forest Environments: An Open-Source Tool}

\author[C. Newlands \& K.P. Zauner]
{\parbox{\textwidth}{\centering Callum Newlands\orcid{0000-0003-4798-9306}
        and Klaus-Peter Zauner
        }
        \\
{\parbox{\textwidth}{\centering Electronics and Computer Science, University of Southampton, UK\\\texttt{\{cn2g18, kpz\}@soton.ac.uk}
       }
}
}

\begin{document}

\teaser{
\begin{minipage}[b]{0.245\linewidth}
 \centering
 \includegraphics[width=\textwidth]{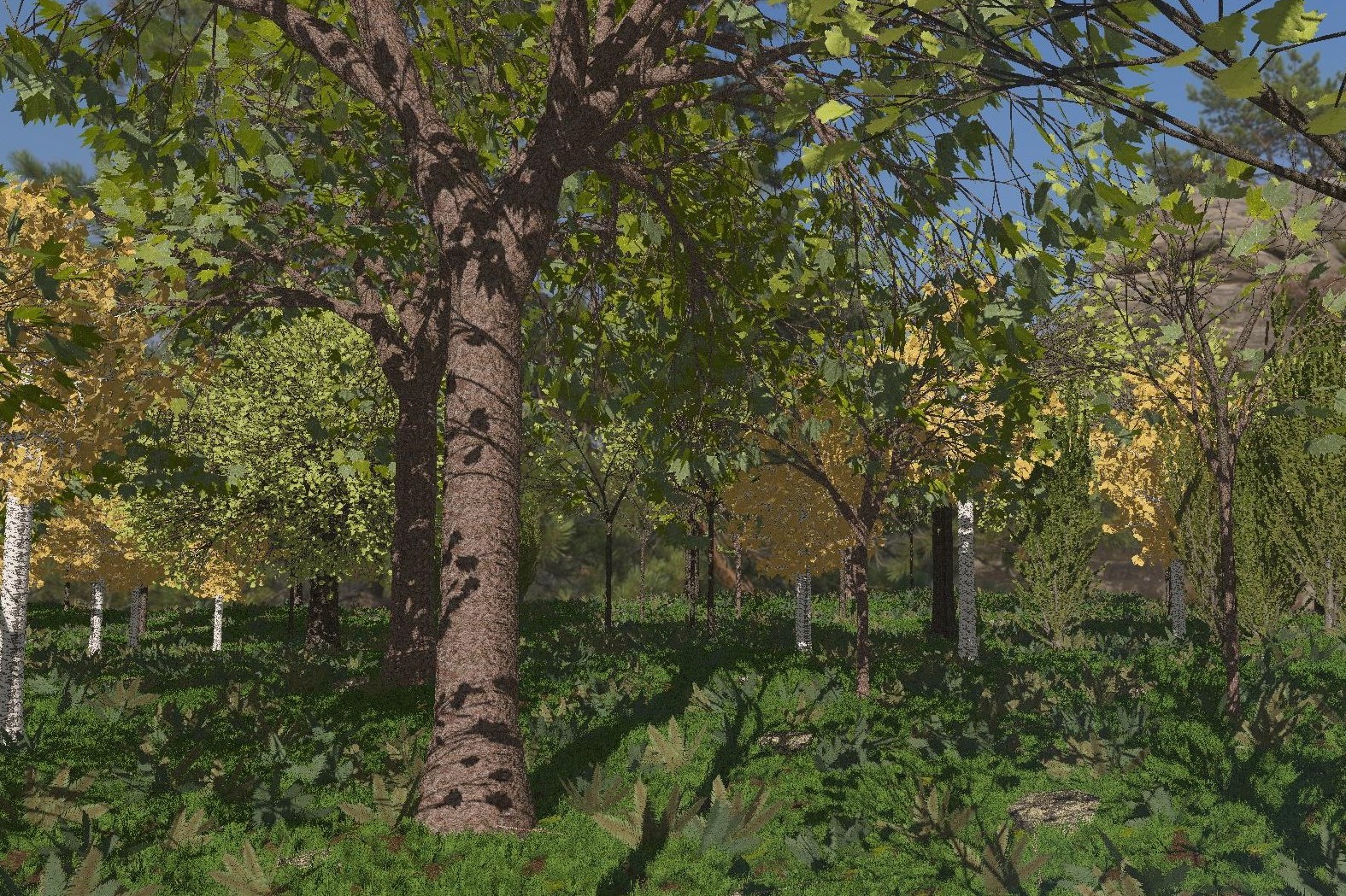}
\end{minipage}
\hfill
\begin{minipage}[b]{0.245\linewidth}
 \centering
 \includegraphics[width=\textwidth]{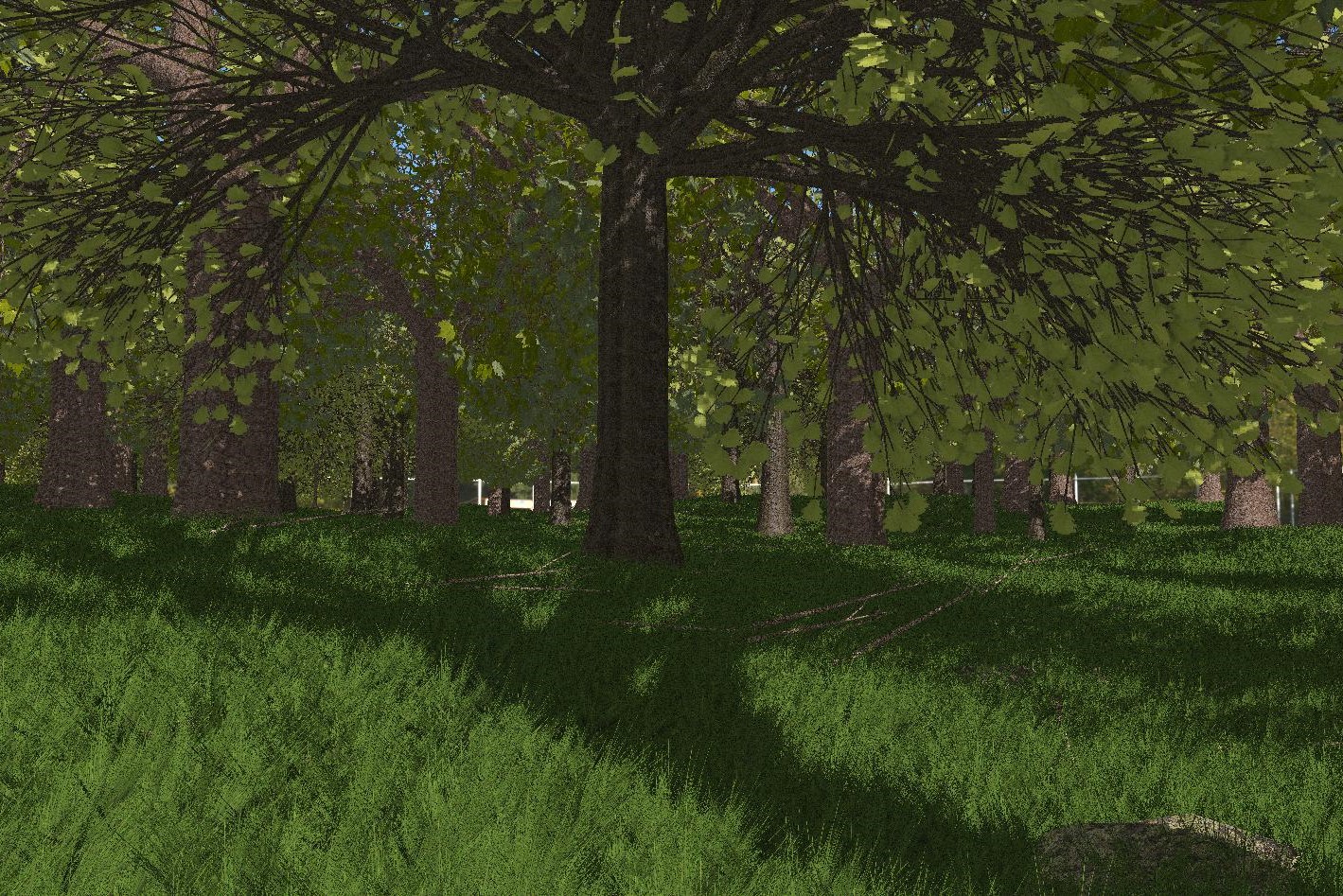}
\end{minipage}
\hfill
\begin{minipage}[b]{0.245\linewidth}
 \centering
 \includegraphics[width=\textwidth]{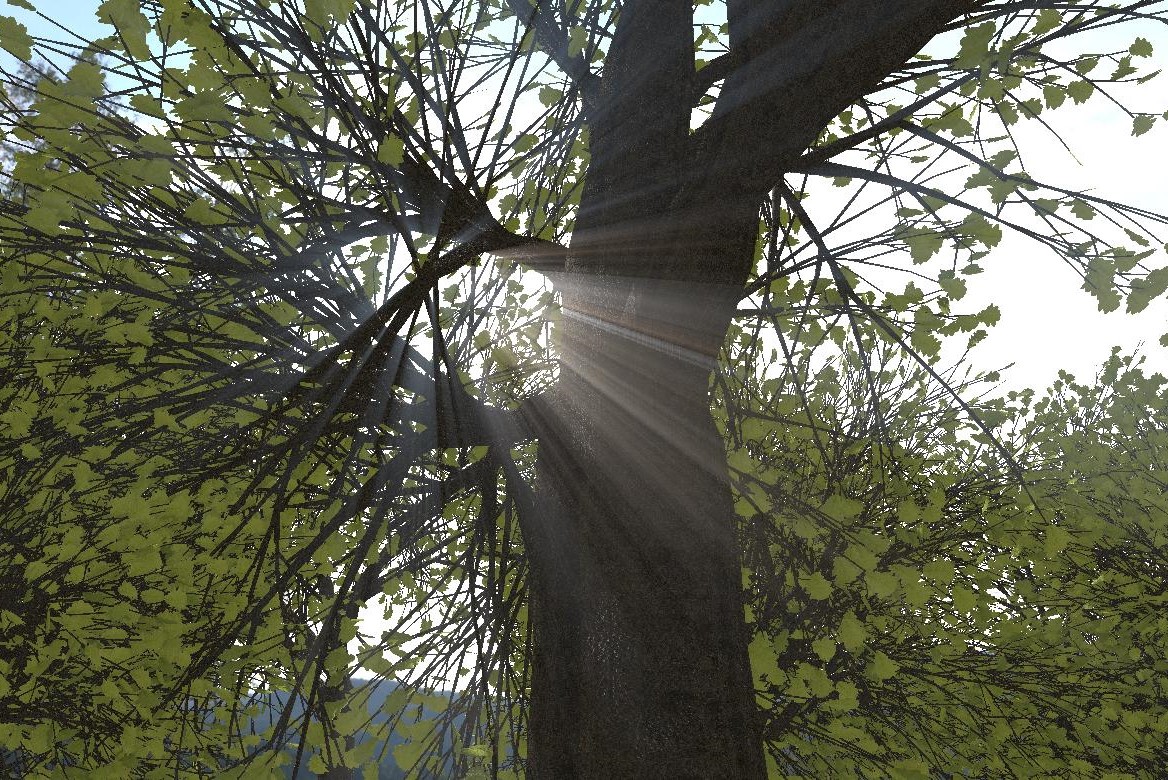}
\end{minipage}
\hfill
\begin{minipage}[b]{0.245\linewidth}
 \centering
 \includegraphics[width=\textwidth]{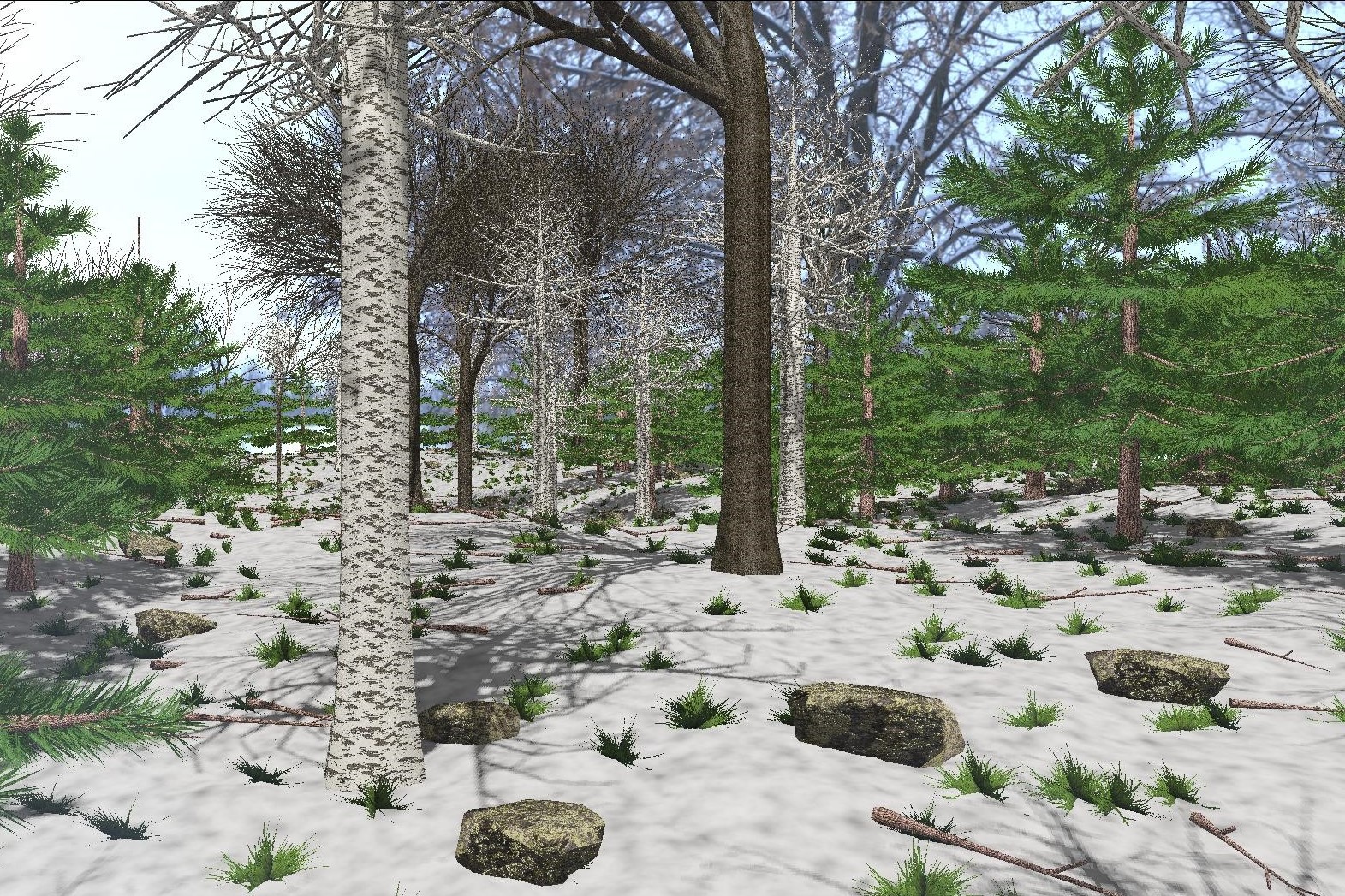}
\end{minipage}
 \centering
  \caption{ Sample image frames of scenes generated by the tool during runs configured with four different parameters sets.  \vspace{4pt}}
\label{fig:teaser}
}


\maketitle
\begin{abstract}
Simulation of forest environments has applications from entertainment and art creation to commercial and scientific modelling. Due to the unique features and lighting in forests, a forest-specific simulator is desirable, however many current forest simulators are proprietary or highly tailored to a particular application. Here we review several areas of procedural generation and rendering specific to forest generation, and utilise this to create a generalised, open-source tool for generating and rendering interactive, realistic forest scenes. The system uses specialised L-systems to generate trees which are distributed using an ecosystem simulation algorithm. The resulting scene is rendered using a deferred rendering pipeline, a Blinn-Phong lighting model with real-time leaf transparency and post-processing lighting effects. The result is a system that achieves a balance between high natural realism and visual appeal, suitable for tasks including training computer vision algorithms for autonomous robots and visual media generation. The application and supporting configuration files can be found at \url{https://github.com/callumnewlands/ForestGenerator}.
\end{abstract}

\section{Introduction}

Procedural generation and efficient rendering of vegetation and natural environments is a growing research area in both the entertainment and commercial industries.
Whilst the former is often concerned with visual attractiveness and generation speed \cite{Morin-16-ArtificialUniverseThat, IDV-20-Speedtree, Landin-20-RoadDungeonsExploring}, the latter often focuses on real-world accuracy and generation flexibility, for purposes such as environmental forecast modelling or simulating real-world locations \cite{House-98-VisualizingRealForest, KomatsuForest-20-KomatsuSimulators, Lenne3D-20-SelectionOurProjects, VirtuellerWald-20-VirtualForest}.

We here present an open-source Forest Generator (available with sample configurations at \cite{NewlandsGitHub}) for procedurally generating and rendering realistic forests which can be virtually navigated in real-time, or through asynchronous connections with external software. The system takes a middle-ground perspective between visual and real-world accuracy, resulting in an approach that is ideal, for example, for training the computer-vision systems of forest navigating robots \cite{Hempe-13-interactiveGroundVegetation, Song-20-FlightmareFlexibleQuadrotor, Tarapore2020SparseRobot}.

The main contributions of this article are two-fold, firstly, a brief review of the existing approaches and literature in several aspects of forest generation and rendering (\sref{Section:Related})---including vegetation generation, rendering optimisation, ecosystem simulations and tree-specific illumination---and secondly, the presentation of a system for generating and rendering realistic forest environments.
\sref{Section:SystemDesign} gives an overview of the system architecture and control flow---which uses 
L-systems for representing and generating a wide range of tree types (\sref{Section:Generation}), along with a tree-specific ecosystem simulation for realistic scene distributions (\sref{Section:GeneratingTreeDistribution}).
The generated scene is rendered using a Blinn-Phong deferred render pipeline, along with a leaf-specific translucency model, and post-processing shaders for shadows and lighting effects (\sref{Section:Rendering}).

\section{Related Work} \label{Section:Related}
Procedural generation and rendering of virtual forest environments comprises many aspects; key among them are plant model generation, vegetation distribution, terrain generation, rendering optimisation, and tree-specific illumination. The following provides a brief overview of the state-of-the-art in these areas.




\subsection{Vegetation} \label{Section:BackgroundVeg}
Methods for vegetation generation can be categorised as procedural and non-procedural.
Non-procedural methods allow for high levels of control and can result in highly detailed models, at the cost of requiring heavy user input---often at a per-plant level.
Key non-procedural methods include manual modelling, and image-based methods such as photogrammetry \cite{Bradley-13-ImageBasedReconstruction,Patel-18-UsingPhotogrammetryMegascans}. 
%
Procedural models are constructed algorithmically and controlled through parameter values without requiring high levels of manual input or expertise.
Several hybrid methods have also been developed to reduce the time taken to create plants under non-procedural methods, by combining algorithmic generation with user descriptions (component graphs) \cite{Lintermann-99-InteractiveModelingPlants} or sketch-guided modelling \cite{Benes-11-GuidedProceduralModeling}.

Early procedural models for trees include the H-model and P-model \cite{Honda-82-TwoGeometricalModels, Prusinkiewicz-90-algorithmicBeauty} which defined branch geometry parametrically in terms of their parent branches through angles and lengths, and further parametric models were subsequently proposed \cite{Weber-95-CreationRenderingRealistic, DiGiacomo-01-interactiveforest, Hewitt-17-ProceduralGenerationTree}. Several methods have been introduced which take into account the fractal-like (recursive) nature of tree generation, and operate in terms of repeated sub-modules including Iterated Function Systems \cite{Barnsley-88-FractalsEverywhere} and self-similar branch modules \cite{Makowski-2019-SyntheticSilviculture}. 

The most well-explored plant modelling technique based on the idea of recursive hierarchies is the concept of L-systems. L-systems are a parallel string-rewriting system and were introduced by Lindenmayer in order to model the growth of cellular organisms \cite{Lindenmayer-68-MathematicalModelsCellular}. There are many types and extensions of L-systems \cite{Prusinkiewicz-86-GraphicalApplicationsL, Prusinkiewicz-90-algorithmicBeauty, Prusinkiewicz-02-fromTheory} including (deterministic) context-free, context-sensitive, stochastic, parametric, differential \cite{Prusinkiewicz-93-AnimationPlantDevelopment}, environmentally-sensitive \cite{Prusinkiewicz-94-topiary} and open \cite{Mech-96-VisualModelsPlants}.  
However, often context-sensitive L-systems, stochastic L-systems, and parametric L-systems are combined, as is the case here.



Prusinkiewicz was the first to propose a graphical interpretation of L-systems using a `turtle' \cite{Prusinkiewicz-86-GraphicalApplicationsL}. After the desired number of iterations of an L-system, the resulting string can be interpreted by the turtle by reading left to right and acting on each symbol in a predefined manner. The turtle interpretation, combined with the expressive power of L-systems, allows for the generation of many complex graphics from the specification of simple rules, and was later extended into 3D and combined with parametric L-systems to allow for rotation and movement by arbitrary amounts \cite{Prusinkiewicz-90-algorithmicBeauty}. A further key extension in the field of plant representations was the inclusion of brackets (\texttt{[}  and \texttt{]}) to allow for branching structures such as trees to be represented \cite{Lindenmayer-68-MathematicalModelsCellular} along with their turtle interpretation through the use of a stack \cite{Prusinkiewicz-86-GraphicalApplicationsL}. 

Another category of procedural plant generation methods are particle-based methods, which simulate the motion of particles to construct the branch structure of trees. These include Diffuse-Limited Aggregation \cite{Witten-81-DiffusionLimitedAggregation}, Space Colonisation  \cite{Runions-07-ModelingTreesSpace} and particle flow modelling \cite{Neubert-07-ApproximateImageBased, Nuic-19-trees}.

As well as the generation of plant models, the distribution of plants within a scene is critical to realism and visual appeal. 
The simplest approach for distributing vegetation is to sample a random probability distribution, or use an algorithm such as Poisson disk sampling \cite{Cook-86-StochasticSamplingComputer}. This na\"{i}ve approach results in unrealistic plant distributions and specialised methods have been proposed. These can be grouped according to whether plant positions are generated from global distribution properties (\textit{global-to-local}) or whether the overall distribution arises from simulating interactions between individual plants (\textit{local-to-global}) \cite{Lane-02-GeneratingSpatialDistributions}.

Several global-to-local methods have been proposed, including: modelling ecotypes from terrain data \cite{Hammes-01-ModelingEcosystemsData}, probability-deformation-kernels \cite{Lane-02-GeneratingSpatialDistributions}, halftoning algorithms \cite{Deussen-98-RealisticModelingRendering}, and sample pattern extrapolation \cite{Xu-15-SampleBasedVegetation}. Alsweis and Deussen presented a method for using Wang Tiles, along with the `field of neighbourhood' (FON) distribution model \cite{Berger-00-NewApproachSpatially}
to model the influence of plants on the growth of other plants in their surrounding area \cite{Alsweis-06-WangTilesSimulation}.
This idea was developed further using Poisson disk distributions (PDDs) and configurable ecosystem properties \cite{Weier-13-GeneratingRenderingLarge}, and to accommodate plants of different sizes without size-related bias \cite{Onrust-17-EcologicallySoundProcedural}. Recently, Nascimento \etal presented a method which takes advantage of GPU parallelism, and uses the idea of ecosystems from \cite{Hammes-01-ModelingEcosystemsData} along with layering, to allow for different sized plants with per-layer parameter maps and cross-layer growth interactions  \cite{Nascimento-18-GpuBasedReal}.

Local-to-global methods typically allow for higher plant-level interactions, at the cost of higher compute times, or possible inconsistencies at a global scale. Early models were based on open L-systems and distributed plants by simulating resource competition or enforcing species-terrain preferences \cite{Deussen-98-RealisticModelingRendering}. This was extended to utilise multiset L-systems and simulate succession and plant-clustering (using deformation kernels) \cite{Lane-02-GeneratingSpatialDistributions}. Bene\v{s} proposed another local-to-global algorithm in the form of a survival simulation on randomly placed plants \cite{Benes-02-StableModelingLarge}. The simulation models the growth and death of plants, seed dispersal and clustering, resource competition and removal of colliding plants. Mikuli\v{c}ic and Mihajlovi\'{c} compared random scattering (with collision detection) to a similar survival simulation and found that the simulation produced a more natural result \cite{Mikulicic-16-ProceduralGenerationMediterranean}.

\subsection{Terrain}
Terrain generation methods are based on one of two representations: \textit{heightmaps}---which store the terrain as a 2D image representing terrain height---and \textit{voxel terrain}---where voxels are elements of a 3D volume grid (typically cubes). The use of heightmaps restricts the terrain to be one continuous surface with no caves or overhangs, however this can be overcome with layered heightmaps \cite{Benes-01-LayeredDataRepresentation} or heightmap transformations \cite{Gamito-03-ProceduralLandscapesOverhangs}. Voxel terrain does not have this limitation but generally has a much higher memory cost and therefore is typically only used in situations where the terrain needs to be edited interactively. Here, only procedural, heightmap-based methods will be discussed, however non-procedural methods include manual modeling, using real-world data, and sketch-based methods \cite{Gain-09-TerrainSketching, Hnaidi-10-FeatureBasedTerrain}.

Terrain height functions map a coordinate pair to a height value to create heightmaps before or during render time. One of the most common categories are noise functions---due to the fact that they can be evaluated without relying on neighbouring points---and include random noise, coherent noise, fractal noise and ridged noise \cite{Olsen-04-terrain, Haggstrom-06-landscapes, Dunn-16-ProceduralGenerationRendering, Freiknecht-17-virtualWorldsSurvey}.

Aside from terrain-height functions, several iterative generation methods have been proposed including midpoint displacement, diamond-square subdivision and faulting algorithms \cite{Belhadj-05-ModelingLandscapesRidges, Haggstrom-06-landscapes, Kahoun-13-library}. At a larger terrain-scale, many physics-based erosion models have also been proposed  \cite{Musgrave-89-SynthesisRenderingEroded, Chiba-98-ErosionModelBased, Benes-01-LayeredDataRepresentation, Benes-02-VisualSimulationHydraulic, Benes-06-HydraulicErosion, Mei-07-FastHydraulicErosion, Kristof-09-HydraulicErosionUsing, Jako-11-FastHydraulicThermal, Kurowski-12-ProceduralGenerationMeandering, Beyer-15-ImplementationMethodHydraulic, Backes-18-RealTimeMassive}, however, due to the comparatively small scale of forests, these methods are somewhat unsuitable for the problem discussed here and are included here only for completeness.

Once the terrain has been generated, post-generation techniques can be applied to increase realism or reduce render time. Three of the most common mesh-optimisation methods are: ROAM \cite{Duchaineau-97-RoamingTerrainReal}, the Lindstrom-Coller Simplification \cite{Lindstrom-96-RealTimeContinuous} and Geometrical MipMapping \cite{Boer-00-FastTerrainRendering}. Terrain smoothing can be performed by applying common imaging filters to the heightmaps \cite{Smelik-09-terrainSurvey}, or using agents which perform random walks and average the height of points \cite{Doran-10-ControlledProceduralTerrain}.

\subsection{Rendering Optimisation} \label{Section:BackgroundRenderingOpt}
Tree models can have in the order of 100,000 polygons \cite{Fuhrmann-05-extremeSimplification} (see supplementary material), therefore
rendering even a few tree models in full detail can quickly become infeasible if the scene is to be navigable with reasonable frame rates. One of the key concepts in rendering complex scenes is Level of Detail (LOD) which involves presenting models using different methods based on their distance from the viewpoint. 

Plant model LOD rendering techniques can be broadly classified into three categories: geometric simplification, point-based, and image-based. Geometric simplification methods reduce the number of polygons in the model and include general polygonal simplification methods (reviewed in \cite{Heckbert-97-SurveyPolygonalSurface}), as well as plant-specific models such as FSA \cite{Remolar-02-GeometricSimplificationFoliage}, BFSA and TFSA \cite{Deng-10-MultiresolutionFoliageForest}. Point-based methods partially or totally replace models with point-cloud representations, as points are typically faster to render than polygon vertices \cite{Reeves-85-ApproximateProbabilisticAlgorithms, Weber-95-CreationRenderingRealistic, Stamminger-01-InteractiveSamplingRendering, Wand-01-RandomizedZBuffer, Deussen-02-interactiveVisualization}. Image-based methods replace 3D model meshes with 2D representations---which have much smaller polygon counts---and result in the highest vertex compression rate at the cost of reduced detail. 

The most common image-based method is a billboard, which replaces a model with a flat texture representation mapped on a rectangular plane.
There are several types of billboard, which reflect the trade-off between complexity reduction and accuracy to the original model \cite{Decoret-03-billboardClouds, Decaudin-04-RenderingForestScenes, Behrendt-05-RealisticRealTime, Dietrich-05-RealisticInteractiveVisualization, Fuhrmann-05-extremeSimplification, Haggstrom-06-landscapes, Deng-10-MultiresolutionFoliageForest, Bao-11-RealisticRealTime, Jakulin-00-InteractiveVegetationRendering, Bao-09-BillboardsTreeSimplification}, with the most notable types being camera-aligned billboards (one billboard per model), crossed billboards (2--3 billboards per model) and billboard clouds (a set of $n$ billboards). 
Among the limitations of 2D billboard representations are the
artefacts when viewed at shallow angles, and the lack of parallax and depth. As a result, several other techniques have been presented which consider model depth. The most notable methods, with regards to walk-through forests, are ``Sprites with Depth'' \cite{Schaufler-97-NailboardsRenderingPrimitive, Shade-98-LayeredDepthImages}, and Layered Depth Images (LDIs) \cite{Shade-98-LayeredDepthImages,Max-95-RenderingTreesPrecomputed, Max-96-HierarchicalRenderingTrees, Max-99-HierarchicalImageBased}; with other methods including volumetric textures \cite{Decaudin-04-RenderingForestScenes}, volumetric billboards \cite{Decaudin-09-VolumetricBillboards}, ``2.5D imposters'' \cite{Szijarto-03-HardwareAcceleratedRendering} and Hierarchies of Bidirectional Texture Functions (HBTs) \cite{Meyer-01-InteractiveRenderingTrees}.

There are several techniques that are used alongside LOD in order to optimise the rendering further or remove visual artefacts (such as ``popping'' between LOD levels); including quadtree data structures \cite{Bruneton-08-RealTimeRendering, Bruneton-12-RealTimeRealistic, Onrust-17-EcologicallySoundProcedural, Kohek-18-InteractiveLargeScale}, instancing \cite{Deussen-98-RealisticModelingRendering, Kenwood-14-efficientForests, Vries-20-LearnOpengl}, GPU mesh construction \cite{Bruneton-12-RealTimeRealistic, Mikulicic-16-ProceduralGenerationMediterranean, Onrust-17-EcologicallySoundProcedural}, importance reduction \cite{Deussen-02-interactiveVisualization}, cross-fading (or alpha-blending) \cite{Jakulin-00-InteractiveVegetationRendering, Haggstrom-06-landscapes}, alpha-to-coverage \cite{Jens-09-responsiveGrass, Kharlamov-07-GpuGems3}, and fizzle LOD \cite{Whatley-05-GpuGems2, Kharlamov-07-GpuGems3}.

\subsection{Tree-Specific Illumination Models}

Forest environments exhibit unique lighting behaviours, in particular, self-shadowing between leaves, and complex indirect lighting due to leaf translucency and light scattering. Accordingly, tree-specific illumination models have been suggested. Qin \etal presented one of the first models, limited to a fixed position and offline (non-real-time) rendering \cite{Qin-03-FastPhotoRealistic}. 
%
%
Both Behrendt \etal and Candussi \etal introduced real-time methods for rendering trees represented as dynamically rendered billboard clouds, with differing lighting models to support either accurate external lighting of large forests \cite{Behrendt-05-RealisticRealTime} or realistic, scalable, real-time animation \cite{Candussi-05-realisticTrees}.

Wang \etal presented a real-time lighting model which accommodated subsurface scattering and transmittance of leaves resulting in translucency \cite{Wang-05-RealTimeRendering}. The model operates on a per-leaf basis and defines spatially-varying bidirectional reflectance distribution functions (BRDFs) and bidirectional transmittance distribution functions (BTDFs) as parameter maps obtained from real leaves. Habel \etal presented a similar model in \cite{Habel-07-PhysicallyBasedReal}, but using an 8D bidirectional scattering-surface reflectance distribution function (BSSRDF) for the back face of the leaves (as opposed to a 4D BTDF). Rather than considering individual leaves, as the previous two models do, Boulanger \etal introduced a lighting model which probabilistically models the spacial distribution of leaves \cite{Boulanger-08-RenderingTreesIndirect}.  The model results in lighting close to Monte-Carlo PBR rendering, however is limited to a uniform distribution of leaf positions and normals, so has limited accuracy for some tree types. 
%
As ray-traced methods become feasible in real-time on consumer hardware,
the focus is shifting towards them \cite{Geist-08-LightingModelFast, Bruneton-12-RealTimeRealistic}. 

\section{System Design} \label{Section:SystemDesign}

\begin{figure}
    \centering
  \includegraphics[width=\linewidth]{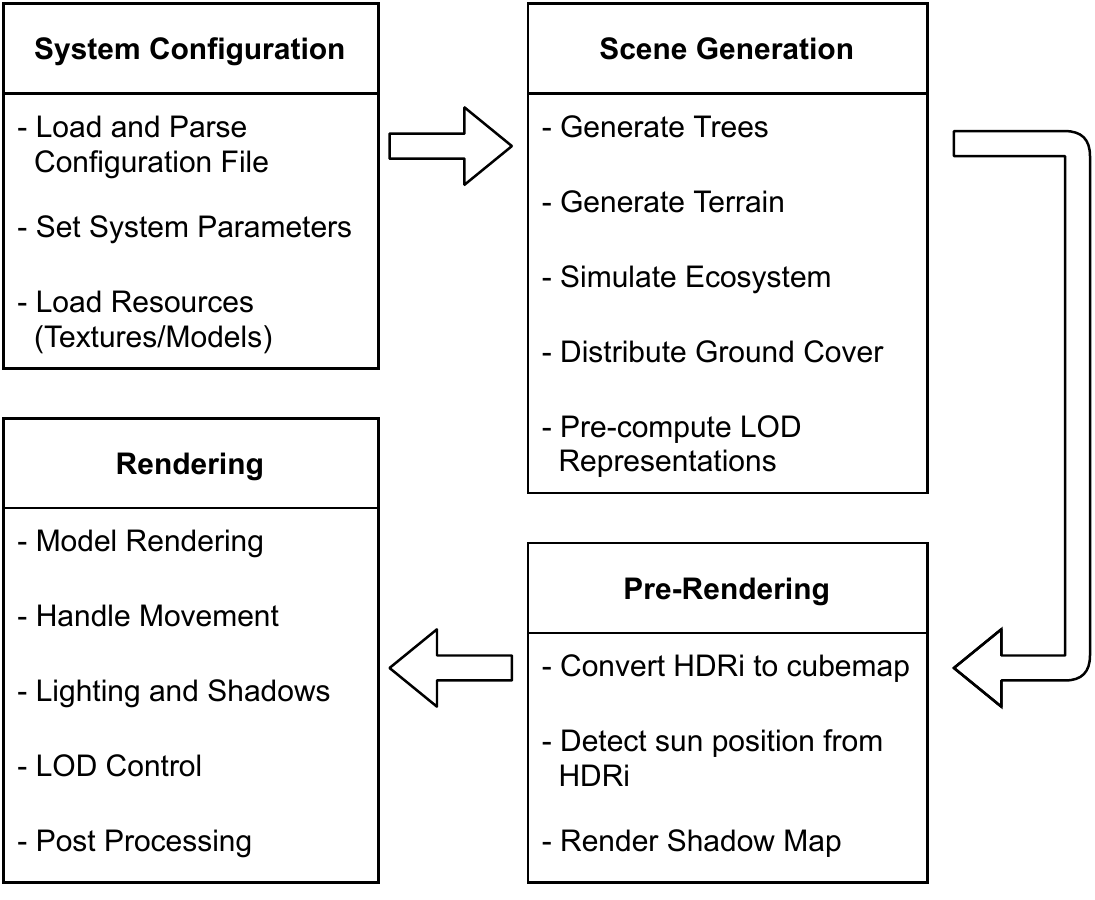}
  \caption[The 4 stage system flow]{The four-stage system flow for generating a forest scene.}
  \label{Figure:system-flow}
\end{figure}

The Forest Generator constructs and renders interactive visualisations of forest scenes in four stages (\fref{Figure:system-flow}).
The generation and rendering are fully configurable through text files (see \cite{NewlandsGitHub}).  These configuration files---in human-readable YAML format---facilitate a wide range of different forest characteristics to be generated, including the creation of new tree models with differing geometry and textures. The seed for the pseudo-random generator can also be specified to allow for identical runs of the application.

Based on the configuration, the application generates the object models and distributes them within the virtual scene. Next, any pre-rendering operations are carried out---such as shadow map rendering---and finally, the system enters the render loop for the final navigable scene.

Movement around the scene at render time can be controlled in two ways: manually and asynchronously. With manual control, the user can pan the view in real-time using the mouse and move around the scene using the keyboard. 
In asynchronous mode, the application blocks after each frame and waits for a new line of input on the standard input stream, to parse and react to, before rendering again. 
As well as real-time output of the scene, the rendered frames can be saved as image files. This, combined with the asynchronous input mode, allows for external software (such as computer vision systems, frame-to-video converters, or image-to-image style transformers) to be easily connected to the input and output of the system, as well as allowing the system to be run in a headless environment. 

\subsection{Scene Construction and Storage} \label{Section:SceneStorage}

The scene generation and pre-rendering stages are performed offline, which limits the size of the forest to being finite and has the drawback of large memory costs---as the entire scene needs to be stored in memory---however, due to the fact that real-world forests are also finite, and the increased realism and quality facilitated by this method, this is deemed to be a worthwhile trade-off. 

To store the generated scene, a quadtree data structure is used. In this case, the tree is complete, as it is fully constructed during the generation phase, with different depth nodes selected to be visible at render time. This enables various parts of the scene to be rendered at different levels of detail (LOD) (\sref{Section:RenderingLOD}) without having to calculate the distance to each object, as well as enabling efficient view-frustum-culling.

The scene is constructed in the following stages:
\begin{enumerate}
    \item The quadtree data structure is constructed recursively to a max depth $D$.
    \item At each leaf node, a terrain tile is constructed (\sref{Section:GeneratingTerrain}) and ground cover for the tile is generated and distributed (\sref{Section:GeneratingGroundCover}).
    \item Minimum and maximum tree masks are calculated and trees are distributed with an ecosystem simulation  (\sref{Section:GeneratingTreeDistribution}).
    \item Models for trees of the required growth stages are constructed (\sref{Section:GeneratingTrees}) and inserted into the relevant leaf nodes.
\end{enumerate}

As all scene models are constructed before render time, this can result in a large memory cost. To minimise this, tree meshes are stored in an object pool and instances are replicated throughout the scene with differing transformations. The number of variants for each growth stage (iteration count) of a tree species can be controlled through a system parameter, to reduce repetition as desired.
Ground cover objects, due to their high per-tile density, are rendered using terrain-tile level GPU instancing. This reduces the overhead of sending geometry to the GPU, whilst still allowing tiles to be culled when out of view (\sref{Section:RenderingLOD}).

\section{Generation} \label{Section:Generation}
Prior to rendering, the system generates the mesh data for each of the elements in the forest scene: trees, terrain and ground cover (plants, twigs, rocks, etc.), and then carries out an ecosystem simulation to distribute the trees within the scene.

\subsection{Trees} \label{Section:GeneratingTrees}
To generate tree models, parametric, stochastic, context-sensitive L-Systems (Section~\ref{Section:BackgroundVeg}) were implemented along with a 3D turtle interpreter. 
The implemented turtle holds its position, heading, and up vectors $\in \mathbb{R}^3$; current radius (to allow for branch tapering); and the vertices of the most recent cross-section. 

%
%
With each interpreted symbol---see supplementary material for the full list of symbols and modules (parameterised symbols)---the turtle's state is transformed using standard 3D matrix transformations, and lists of mesh vertices and sub-model references (position, heading, up-vector and sub-model list index) are added to where necessary by the interpreter. If tropism is enabled---by seeing a module of the form $T(v_1, v_2, v_3, e)$---each time the cross-section is transformed, a (small) rotation (with rotation factor $\alpha$) is applied in the direction $\bm v = [v_1, v_2, v_3]^T$ in the same manner as \cite{Prusinkiewicz-90-algorithmicBeauty} to simulate the growth of plants dependent on the direction of external forces (such as gravity or sunlight):
\begin{equation}
        \alpha = \frac{\theta_e}{\theta_v} = \frac{e \, \norm{\bm h \times \bm v}}{\arccos\left(\frac{\bm h \cdot \bm v}{\norm{\bm h} \, \norm{\bm v}}\right)}
\end{equation}
where $\theta_e$ is the angle of rotation for the tropism, proportional to the elasticity ($e$) of the branch, and $\theta_v$ is the angle between the turtle heading $\bm h$ and tropism direction $\bm v$.

\begin{figure}[b!]
    \centering
    \tdplotsetmaincoords{70}{0}
    \begin{tikzpicture}[tdplot_main_coords, scale=0.7]
    \def\RI{2}
    \def\RII{1.25}
    \def\RIII{2}
    \def\RIV{1.25}
    
    \draw[thick] (\RI,0)
      \foreach \x in {0,300,240,180} { --  (\x:\RI) node at (\x:\RI) (R1-\x) {} };
    \draw[dashed,thick,opacity=0.7] (R1-0.center)
      \foreach \x in {60,120,180} { --  (\x:\RI) node at (\x:\RI) (R1-\x) {} };
      
    \foreach \x in {0,60,120,180,240,300} { \draw[dashed,opacity=0.7] (R1-\x.center)--(0,0); };
    \fill (240:\RI) (R1-240) circle (.5ex) node[below,anchor=north west] {B};
    \fill (300:\RI) (R1-300) circle (.5ex) node[below,anchor=north east] {A};
    
    \begin{scope}[yshift=2cm]
        \draw[thick] (\RII,0)
          \foreach \x in {0,300,240,180} { --  (\x:\RII) node at (\x:\RII) (R2-\x) {} };
        \draw[dashed,thick,opacity=0.7] (R2-0.center)
          \foreach \x in {60,120,180} { --  (\x:\RII) node at (\x:\RII) (R2-\x) {} };
        \foreach \x in {0,60,120,180,240,300} { \draw[dashed,opacity=0.7] (R2-\x.center)--(0,0); };
        \fill (240:\RII) (R2-240) circle (.5ex) node[below,anchor=north west] {D};
        \fill (300:\RII) (R2-300) circle (.5ex) node[below,anchor=north east] {C};
    \end{scope}
    \foreach \x in {0,180,240,300} { \draw[thick] (R1-\x.center)--(R2-\x.center); };
    \foreach \x in {60,120} { \draw[dashed,thick,opacity=0.7] (R1-\x.center)--(R2-\x.center); };

    \begin{scope}[yshift=-2cm]
    \path[fill=gray!30] (\RIII,0)
          \foreach \x in {0,60,120,180,240,300} { --  (\x:\RIII) node at (\x:\RIII) (R3-\x) {}};  
        \draw[thick] (\RIII,0)
          \foreach \x in {0,300,240,180} { --  (\x:\RIII) node at (\x:\RIII) (R3-\x) {} };
        \draw[dashed,thick,opacity=0.7] (R3-0.center)
          \foreach \x in {60,120,180} { --  (\x:\RIII) node at (\x:\RIII) (R3-\x) {} };
        \foreach \x in {0,60,120,180,240,300} { \draw[dashed,opacity=0.7] (R3-\x.center)--(0,0); };
    \end{scope}
    \foreach \x in {0,180,240,300} { \draw[thick] (R3-\x.center)--(R1-\x.center); };
    \foreach \x in {60,120} { \draw[dashed,thick,opacity=0.7] (R3-\x.center)--(R1-\x.center); };
    
    \begin{scope}[yshift=4cm]
        \path[fill=gray!30,opacity=0.3] (\RIV,0)
          \foreach \x in {0,60,120,180,240,300} { --  (\x:\RIV) node at (\x:\RIV) (R4-\x) {}};  
        \draw[thick] (\RIV,0)
          \foreach \x in {0,60,120,180,240,300,360}
            { --  (\x:\RIV) node at (\x:\RIV) (R4-\x) {}};
          
        \foreach \x in {0,60,120,180,240,300} { \draw[] (R4-\x.center)--(0,0); };
    \end{scope}
    
    \foreach \x in {0,180,240,300} { \draw[thick] (R2-\x.center)--(R4-\x.center); };
    \foreach \x in {60,120} { \draw[dashed,thick,opacity=0.7] (R2-\x.center)--(R4-\x.center); };
    \end{tikzpicture}

    \caption[Diagram of the geometry generated by a turtle interpreting the commands $F(1)!(0.6)F(1)F(1)$]{Diagram of the geometry generated by a turtle interpreting the commands $F(1)!(0.6)F(1)F(1)$, with $N$ (number of edges) $=6$. Vertices $A, B, C, D$ outline a connecting face between 2 cross-sections.}
    \label{fig:branch-mesh}
\end{figure}
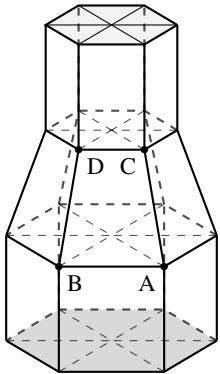

To construct the branch geometry, adjacent cross-sections of $N$ edges are considered pairwise. For each adjacent pair of vertices in the two cross-sections, a connecting face $ABCD$ is constructed (see \fref{fig:branch-mesh}),
where the normal vector for a vertex is determined using the cross product of its two adjacent edges; (e.g. $\widehat{\bm{n_B}} = \vv{BD} \times \vv{BA}$). Since the faces are flat, the tangent vector for each vertex is simply the vector along one of its adjacent edges. After all connecting faces have been created, the four normal vectors at each vertex are averaged to create smooth normals across edges, and similarly for the tangents. This prevents flat shading and allows for a lower value of $N$ to be used with the same visual appearance. Each quadrilateral face is divided into two primitive triangles $ABD$ and $ADC$ for rendering. To save memory, the faces of the interior cross-sections are not generated. 

To inject a sub-model (such as a leaf or pine cone) into the mesh, given a sub-model reference (position $\bm p$, heading $\bm h$, up-vector $\bm u$ and model index $i$), the geometry is first rotated so that $\bm x = [1, 0, 0]^T$ aligns with $\bm h$, and then rotated around $\bm y = [0, 1, 0]^T$ with angle 
$\theta = \atantwo \left((\bm y \times \bm u) \cdot \widehat{\bm x},\, \bm y \cdot \bm u \right)$,
so that the $y$-axis of the sub-model aligns with $\bm u$. 
Next, the geometry is translated so that the origin of the sub-model lies at $\bm p$, and the resulting vertex data is added to the combined mesh for all injections of model $i$.

The module $F(d, n_l, m_i, \alpha_r, \alpha_l)$ is used to inject leaves which grow out of the branches (such as pine needles or poplar leaves) as sub-models. This causes the turtle to distribute $n_l$ copies of sub-model $m_i$ along a spiral around the constructed prism, controlled by the radial angle $\alpha_r$ and lift angle $\alpha_l$. 
To introduce some pseudo-random variation into the model placement, the radial angle ($\alpha_r$) and distance between consecutive sub-models ($s$) are not strictly constant, but instead are offset with a value sampled from a uniform distribution with range $\pm$ 50\% of their (constant) values; resulting in $\alpha_r \in \left[-\frac{\alpha_r}{2}, \frac{\alpha_r}{2} \right)$ and $s \in \left[-\frac{s}{2}, \frac{s}{2} \right)$.

The low LOD representations of the branches are constructed using the same L-system and parameter values, but using a cross-section with 2 edges to create a flat version of the tree with the same branches. For the leaves, adjacent pairs of sub-model references are merged with an average position and angle, and replaced with a model of double width and 1.1 times height in a fashion similar to the ideas presented in \cite{Remolar-02-GeometricSimplificationFoliage}.

\begin{figure}
    \centering
    \begin{subfigure}[c]{0.47\linewidth}
        \centering
        \includegraphics[width=\textwidth]{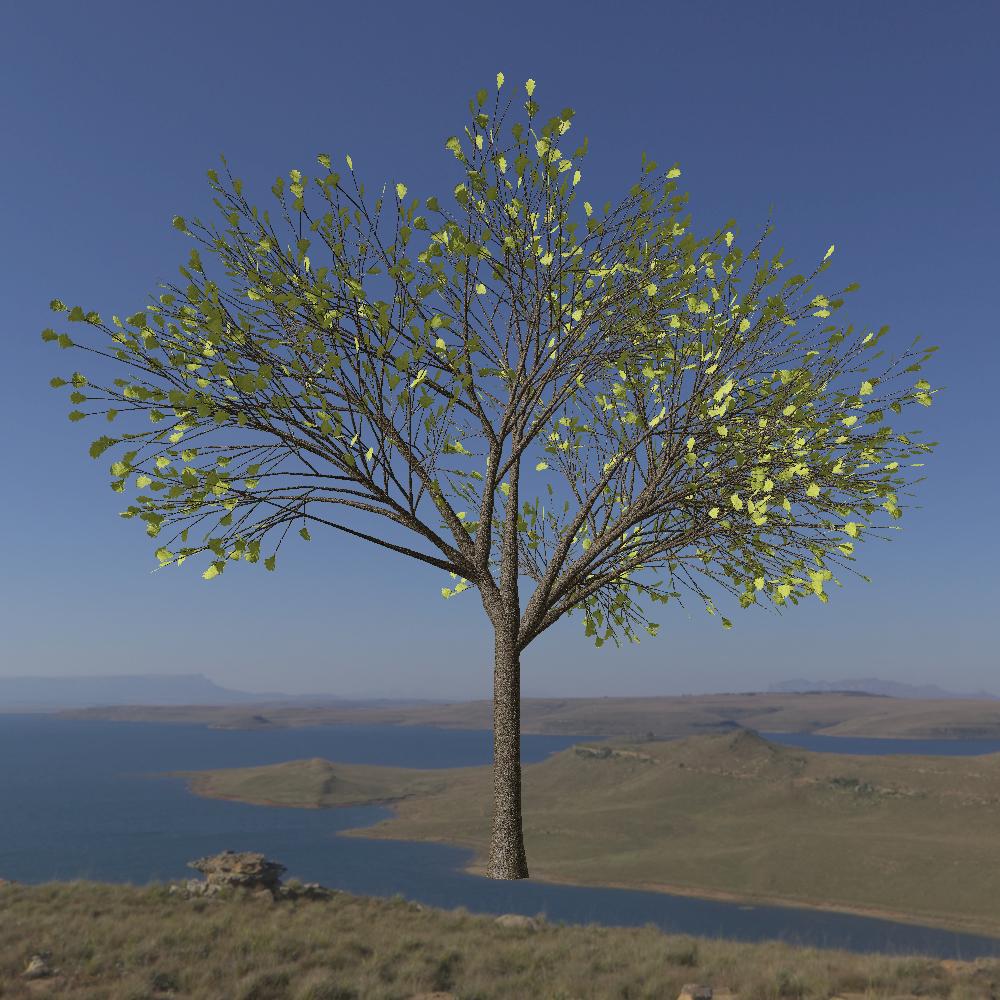}
        \label{subfig:trees-branching}
    \end{subfigure}
    \begin{subfigure}[c]{0.47\linewidth}
        \centering
        \includegraphics[width=\textwidth]{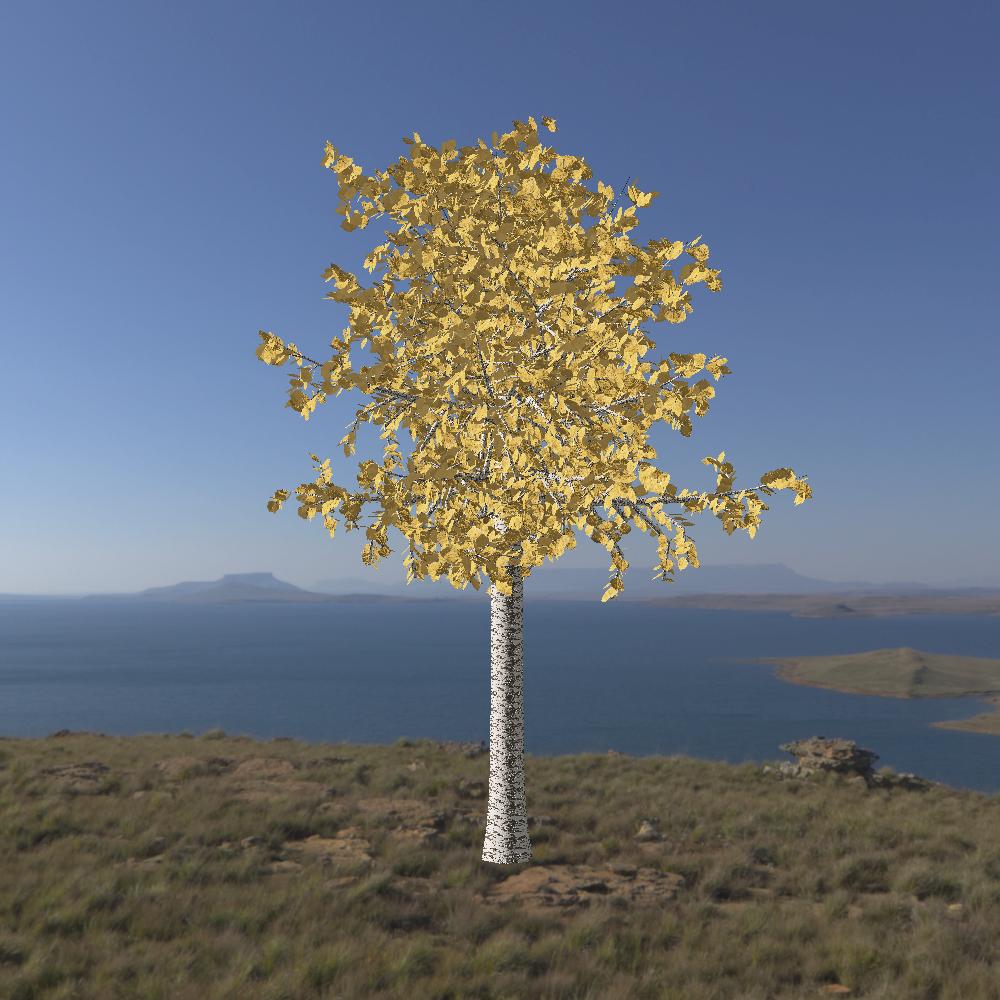}
        \label{subfig:trees-monopodial}
    \end{subfigure}
    \caption[Examples of a branching tree and monopodial tree]{Examples of a branching tree (left) and monopodial tree (right) generated by the system.}
    \label{fig:tree-types}
\end{figure}

A \textbf{branching tree}---in this work---is one where the main trunk (and subsequent branches) split into $n$ branches, and the original trunk ceases. As well as this, there is the possibility for `side branches' to protrude from the trunk or branch before it splits (\fref{fig:tree-types}).

In the L-systems used, a branching tree species is specified with a number of parameters---such as initial trunk width or branch offsets---and a set of $n$ (possible) branchings $Br$. The $i$\textsuperscript{th} branching ($Br_i$) is a branching into $m_i + 1$ sub branches with angles $\theta_i^0,\, \theta_i^1,\, \dots,\, \theta_i^{m_i}$ and probability $p_i$ such that $\sum_i p_i = 1$. $\theta^k$ is the angle around the trunk between branches $k$ and $k+1$.
\begin{equation}
    Br = \left\{\,
        \left\langle\theta_0^0,\, \dots,\, \theta_0^{m_0} ,\, p_0\right\rangle,\, 
        \dots,\, 
        \left\langle\theta_n^0,\, \dots,\, \theta_n^{m_n} ,\, p_0\right\rangle\,
    \right\}
\end{equation}

Pseudo-random variation is achieved by providing upper and lower bounds ($x_{UB}$ and $x_{LB}$ respectively) for the values of each parameter (L-system constant), and then, at each usage, a uniformly distributed random number is sampled from the distribution $x \sim \mathcal{U}\left[x_{LB}, x_{UB}\right)$.

A \textbf{monopodial tree} is one with a single central trunk which extends to the top of the tree and has side branches protruding outwards from it (\fref{fig:tree-types}). Two L-systems are used to accommodate two different types of `side branching': alternating and pine-style. Alternating branches appear in a spiral around the main trunk, with each side branch having a number of \nth{3}-level side branches. Pine-style branches do not appear in a spiral, but rather all appear around the trunk in discrete layers. The side branches themselves are also different in that, at each iteration, they split from a single point into two higher-level side branches.

\subsection{Terrain and Ground Cover} \label{Section:GeneratingTerrain}  \label{Section:GeneratingGroundCover}
The scene terrain is constructed from a heightmap which is generated using 2D fractal simplex noise \cite{Perlin-02-NoiseHardware}
and is configurable through the system parameters. Details of the terrain mesh construction can be found in the supplementary material for this paper.
%

The system has four categories of ground cover: fallen leaves, twigs, crossed billboards (plants) and external models---with varying levels of customisation.
Fallen leaves are generated by positioning leaf sub-models on (just above) the ground mesh with controllable density and scale, and twigs are similarly configurable: generated using a simplified form of the alternating branch monopodial tree L-system (see supplementary material for the specific L-system). Their geometry is formed in the same manner as the trees, and then rotated $\frac{\pi}{2}$ radians around the $x$-axis so the main axis of the twig lies parallel to the floor. 

\begin{figure}
     \centering
     \begin{subfigure}[c]{0.4\linewidth}
         \centering
         \includegraphics[width=\textwidth]{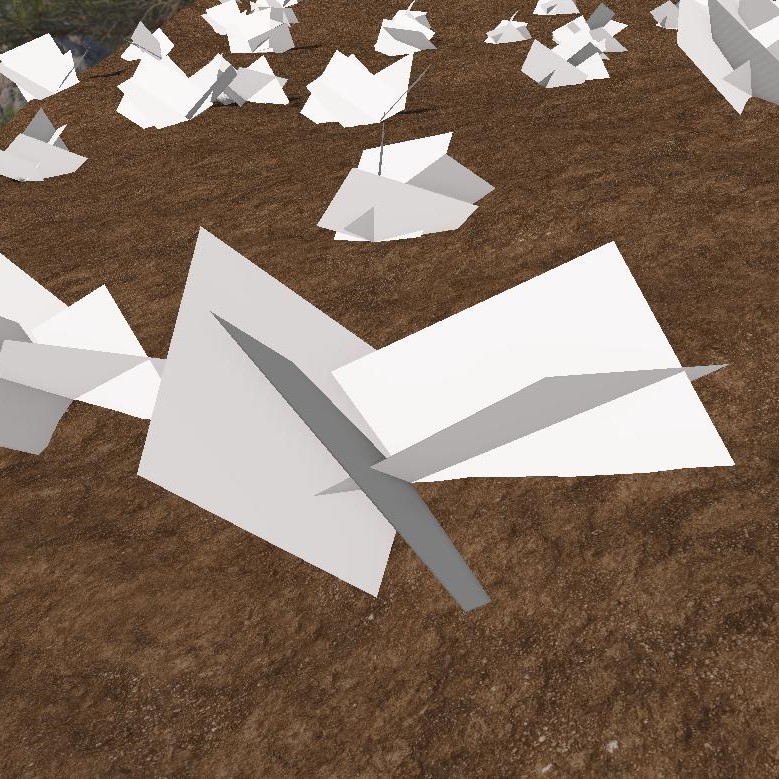}
     \end{subfigure}
     \quad
     \begin{subfigure}[c]{0.4\linewidth}
         \centering
         \includegraphics[width=\textwidth]{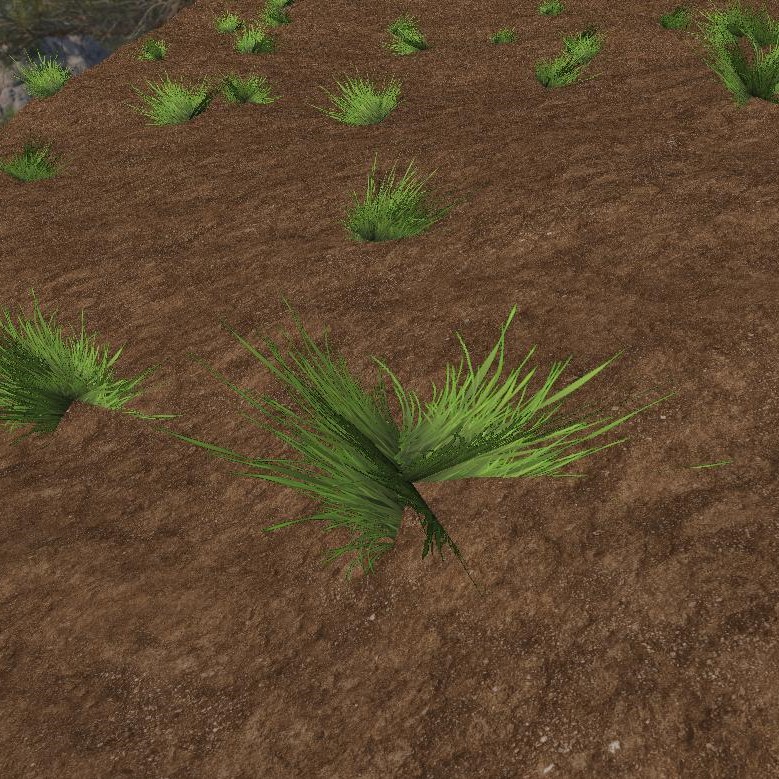}
     \end{subfigure}
     \caption[Example of the board rotations for high LOD crossed billboard objects with four boards and low LOD representations with two boards]{Example of the board rotations for high LOD crossed billboard objects with four boards (left) and their appearance when rendered (right).
     }
     \label{fig:billboard-planes}
 \end{figure}

Crossed billboards (\sref{Section:BackgroundRenderingOpt}) are rendered as $N$ intersecting planes (boards). By default, the system generates grass and two fern types as crossed billboard objects.
Each of the boards is scaled in height and width, and rotated by a random angle $\phi \in \left[\frac{4\pi}{24},\,\frac{10\pi}{24} \right)$ around a random vector $\bm r$ which lies on the $xz$-plane. This orientates the board at a random angle whilst keeping the base of the board at the origin. The board is then rotated by $\theta_i$ radians around the y axis where
\begin{equation}
    \theta_i = \frac{i \pi}{N}, \quad  i \in \left\{\, 0,\, \dots,\, N-1 \,\right\}
\end{equation}
 to distribute the boards around a circle with random angles (\fref{fig:billboard-planes}).
The low LOD representations are constructed using two perpendicular boards.

As well as the three built-in ground cover object types, the system also allows for any external 3D models to be imported and distributed on the ground. For example more-detailed leaf piles or fallen tree logs could be imported from files in common 3D file types. The models (and textures) are loaded using the Open Asset Import (Assimp) Library \cite{LWJGL}, and converted into the local system format at load time.


As mentioned in \sref{Section:SceneStorage}, ground cover objects are drawn with OpenGL instanced rendering. Each instance is given a unique scale, position and rotation in the form of a model matrix. 
The density parameter of each object type ($\rho$) controls the total number of instances ($n_I$) to create:
\begin{equation}
    n_I = \frac{d_0 \cdot \rho}{n_T}w^2 
    \label{eq:objectDensity}
\end{equation}
where $w$ is the width of the scene, $d_0$ is the default density for the object and $n_T$ is the number of `species' of that object type. The division by $n_T$ ensures that the object type remains at a constant density regardless of the value of $n_T$. 
For fallen leaves and twigs, $n_T = 1$.

\subsection{Ecosystem Simulation} \label{Section:GeneratingTreeDistribution}
To distribute the trees throughout the scene, an ecosystem simulation was implemented. This creates a more realistic distribution than random placement, as natural behaviours emerge such as clustering of species and areas of negative growth around trees. The simulation is similar to the ones presented in \cite{Benes-02-StableModelingLarge} and \cite{Mikulicic-16-ProceduralGenerationMediterranean}, but has been adapted specifically for trees. Rather than considering plants as circles on a plane, here, the trees are considered as 3D masks with a trunk cylinder and a canopy cylinder,
which allows for shorter trees to exist underneath the canopy of larger ones, and enables the distinction between branches intersecting tree trunks and intersecting other branches.

To initialise the simulation, a number of plants 
for each specified tree species (determined by Equation~\ref{eq:objectDensity}) are created with random ages $a_i \in \left[0 ,\, a_M \right)$---where $a_M$ is the maximum age for the species---and are randomly placed in the scene largest to smallest.

After initialisation, the simulation runs for $N$ steps, with $N_Y$ steps constituting a year. For each step, the following occur:
\begin{enumerate}
    \item If it is the end of the year, all trees seed a number of new plants in a ring around themselves.
    \item For each pair of colliding trees, the plant with lower viability is removed.
    \item Plants which are older than their maximum age are considered dead and are removed.
    \item Each plant grows (its age is increased by 1).
\end{enumerate}

The viability ($v$) of each plant is determined similarly to \cite{Benes-02-StableModelingLarge} with an age component and a negative feedback component to prevent exponential growth of a plant species and simulate environmental feedback. However, here, the age component threshold ($a_T$) is not fixed at 50\% of the plants age, but is instead variable with a system parameter.  As well as this, a positive feedback component was added to increase the viability of plants with a larger radius,
as the initial implementation was found to lead to the decline (and often species eradication) of larger radius trees. This is likely due to the fact that trees with larger canopies are more likely to collide with another plant, however, in nature, larger trees are often hardier than smaller ones. 

After the simulation has undergone $N$ iterations, a given plant's mesh is generated with $n_i$ iterations of the L-system:
\begin{equation}
    n_i  = \lfloor \widehat{a} \, (i_{\text{max}} + 1 - i_{\text{min}}) \rfloor
\end{equation}
where $i_{\text{max}}$ and $i_{\text{min}}$ are the maximum and minimum number of L-system iterations for the plant's species, and  $\widehat{a}$ is the normalised age for the plant:
\begin{equation}
    \widehat{a} = \text{min}\left( \frac{a}{a_M},\, 1 \right),
\end{equation}
where $a$ is the plant's current agent and $a_M$ is the maximum age for the plant's species.
Next a mesh scale factor $s$ is used to smoothly interpolate between iteration counts:
\begin{equation}
    s  = \text{lerp}(s_{\text{min}},\; s_{\text{max}},\; \widehat{a} \; (i_{\text{max}} + 1 - i_{\text{min}}) - n_i)
\end{equation}
in which $s_{\text{max}}$ and $s_{\text{min}}$ are the upper and lower bounds on the scaling factor for the given plant's species and $\text{lerp}$ is the standard linear interpolation function.
Finally, a random rotation around the vertical axis is applied---as in \sref{Section:GeneratingGroundCover}---to vary the orientations of trees and prevent the appearance of repetition across model instances.

\section{Rendering} \label{Section:Rendering}
To convert the scene meshes into screen-space pixel values, the system carries out six render passes: one shadow pre-pass before the main render loop, one geometry pass, one lighting pass and three post-processing passes. The system uses a deferred rendering pipeline: framebuffers store the result of each pass to be used later. Therefore expensive lighting calculations are only performed for visible pixels. This also enables pixel-level post-processing effects such as screen-space ambient occlusion (SSAO) and volumetric light scattering (\sref{SubSection:PostProcessing}).

The geometry pass uses a number of vertex and fragment shaders for the different object types. The vertex shaders convert the local mesh vertex data into a global vertex position $\bm{p}^g$, a TBN (tangent-bitangent-normal) matrix $\bm T$ and a re-normalised surface normal $\widetilde{\bm{n}}$ \cite{Vries-20-LearnOpengl}. Normal mapping is then carried out in the fragment shaders by sampling fragments from the texture map, which are then multiplied by $\bm T$ to convert from tangent space to world space. 
For fragments which are part of a billboard, the map normal ($\widehat{\bm{n^m}}$) is negated if the `back' face of the board (the face that the board normal ($\widetilde{\bm{n}}$) is facing into) is being viewed---to create a board with separate front and back faces:
\begin{equation}
    \widehat{\bm{n^m}} = 
    \begin{cases}
        -\widehat{\bm{n^m}} ,& \text{if } \widetilde{\bm{n}} \cdot \widehat{\bm{v}} < 0 \\
        \widehat{\bm{n^m}}  ,& \text{otherwise}
    \end{cases}
    \label{eq:billboardNormal}
\end{equation}
in which $\widehat{\bm{v}}$ is the vector from the fragment position to the camera position. Finally, alpha-channel fragment discarding is carried out to enable transparency, and the position, normal, colour and occlusion values are written to framebuffers.
 
As well as rendering the scene models, the geometry pass also renders an HDRI skybox with a `clip-space to normalised-device-space' coordinate transformation to display the skybox at infinite distance \cite{Vries-20-LearnOpengl}. Before the main render loop, the system converts the equirectangular HDRI projection into six 2D textures for the skybox faces.


\subsection{Lighting}

In the lighting pass, a Blinn-Phong lighting model \cite{Blinn-77-ModelsLightReflection} is used with percentage-closer-filtering (PCF) shadow mapping \cite{Reeves-87-RenderingAntialiasedShadows}, and a specific lighting model with translucency for leaves. 
The only light source in the scene is the sun, which is modelled as a directional light with position $\bm p_S$ and colour (strength) $\bm c_S$.

Whilst the sun position ($\bm p_S$) can be set manually by a system parameter, to enable easy changing of light conditions, the position can also be estimated from the HDRi skybox image. The position is estimated by scanning over each pixel in the HDRi texture and summing the RGB values for each pixel in a $3\times3$ kernel around the pixel. The pixel with the largest sum value is then converted into a Cartesian coordinate and scaled to a suitable distance from the scene to give the sun's position vector.

From the sun, the reflected light $L_o$ at a point $\bm x$ in direction $\bm \omega_0$ is given by:
\begin{equation}
    L_o(\bm x,\, \bm \omega_0) = L_a(\bm x) + \left(1-s\left(\bm x,\, \bm \omega_0\right)\right)\left(L_d\left(\bm x\right) + L_s\left(\bm x,\, \bm \omega_0\right)\right)
    \label{eq:blinnPhong}
\end{equation}
where $s$ is the shadow coefficient (1 in full shadow, 0 in full light),
%
$L_d$ is the standard Phong diffuse component \cite{Phong-75-IlluminationComputerGenerated},
%
$L_s$ is the standard Blinn-Phong specular (glossy) component \cite{Blinn-77-ModelsLightReflection},
and  $L_a$ is an ambient component used to approximate indirect lighting, 
\begin{equation}
     L_a(\bm x) = \alpha L_{ao}(\bm x) \bm c_d
\end{equation}
in which $\bm c_d$ is the diffuse colour of the fragment, $\alpha$ is a constant ambient factor and $L_{ao}$ is the screen space ambient occlusion term (\sref{SubSection:PostProcessing}).

A single shadow map is generated for the whole scene, in a pre-render pass, where the scene is rendered with an orthogonal projection matrix $\bm{P_s}$ which defines a view cuboid with width (and height) $w$, near plane distance $d_{near} = 0.1$ and far plane distance $d_{far}$: 
\begin{equation}
    w = \text{max}\left( w_t \sqrt{2} ,\, 50\right)
\end{equation}
\begin{equation}
    d_{far} = \max_{\bm c \in \bm C} \left(\lvert\lvert \bm p_s - \bm c \rvert \rvert \right) + 10
\end{equation}
where $w_t$ is the terrain width, $\bm p_s$ is the sun position and $\bm C$ is the set of corners of the terrain:
\begin{equation}
\begin{aligned}
    \bm C = \biggl\{ \,
    \frac{1}{2}\begin{bmatrix}
        -w_t ,& 0 ,& -w_t
    \end{bmatrix}^T &,\quad
    \frac{1}{2}\begin{bmatrix}
        -w_t ,& 0 ,& w_t
    \end{bmatrix}^T ,\, \\
    \frac{1}{2}\begin{bmatrix}
        w_t ,& 0 ,& -w_t
    \end{bmatrix}^T &,\quad
    \frac{1}{2}\begin{bmatrix}
        w_t ,& 0 ,& w_t
    \end{bmatrix}^T\,
    \biggr\}
\end{aligned}
\end{equation}
These values of $w$ and $d_{far}$ ensure that the entire scene is captured within the view cuboid of the sun, and an orthogonal projection matrix is used to simulate parallel rays of light from the sun. Along with $\bm{P_s}$, a view matrix $\bm{V_s}$ is used to orientate the scene from the sun's position.

The shadow map is then used during the lighting pass to determine if a fragment is in shadow or not ($s(\bm x)$), and PCF is carried out by averaging the shadow coefficients for the 9 pixels of the shadow map in the $3\times3$ kernel around the current fragment.

The lighting model for leaves includes the transparency model proposed in \cite{Habel-07-PhysicallyBasedReal} used alongside the Blinn-Phong shading model. Each leaf is modelled as a two-sided plane (using normal negation as in Equation~\ref{eq:billboardNormal}), and has four textures associated with each side (front and back):
\begin{enumerate}
    \item An albedo (diffuse) colour texture.
    \item A translucency colour texture which holds the colour of translucent light through the leaf.
    \item A normal map.
    \item A half-life basis map which holds the (pre-computed) translucency factor for each point in the leaf represented in the HL2 basis \cite{McTaggart-04-HalfLife2} created with the method proposed in \cite{Habel-07-PhysicallyBasedReal}.
\end{enumerate}

The resulting light from a leaf fragment is given by the following equation (adapted from \cite{Habel-07-PhysicallyBasedReal} for a Blinn-Phong pipeline):
\begin{equation}
    L_{\text{o-leaf}}(\bm x,\, \bm \omega_0) = L_o(\bm x,\, \bm \omega_0) + \beta \bm c_t \odot \bm c_S  \sqrt{\frac{3}{2\pi}}\sum_{i=1}^3 [\bm h_l]_i (\bm H_i \cdot \bm \omega_S)
\end{equation}
where $L_o$ is the standard lighting model for the system (Equation~\ref{eq:blinnPhong}) , $\bm h_l$ is the half-life basis map value, $\bm c_t$ is the translucency colour map value, $\beta$ is a global translucency factor scalar, and $\bm H_i,\, i \in \{1\dots3\}$ are the HL2 basis vectors \cite{Habel-07-PhysicallyBasedReal}.

\subsection{Post-processing} \label{SubSection:PostProcessing}

The system uses the implementation of screen-space ambient occlusion (SSAO) presented in \cite{Vries-20-LearnOpengl} where the post-processing shader operates on fragment depth values and samples a number of values in a normal-aligned hemisphere around each fragment to determine how exposed the point is to indirect lighting, giving an ambient occlusion term $L_{ao}$. A second shader is then used to blur the SSAO texture to reduce noise artefacts.

\begin{figure}[b!]
    \centering
    \begin{subfigure}[c]{0.495\linewidth}
        \centering
        \includegraphics[width=\textwidth]{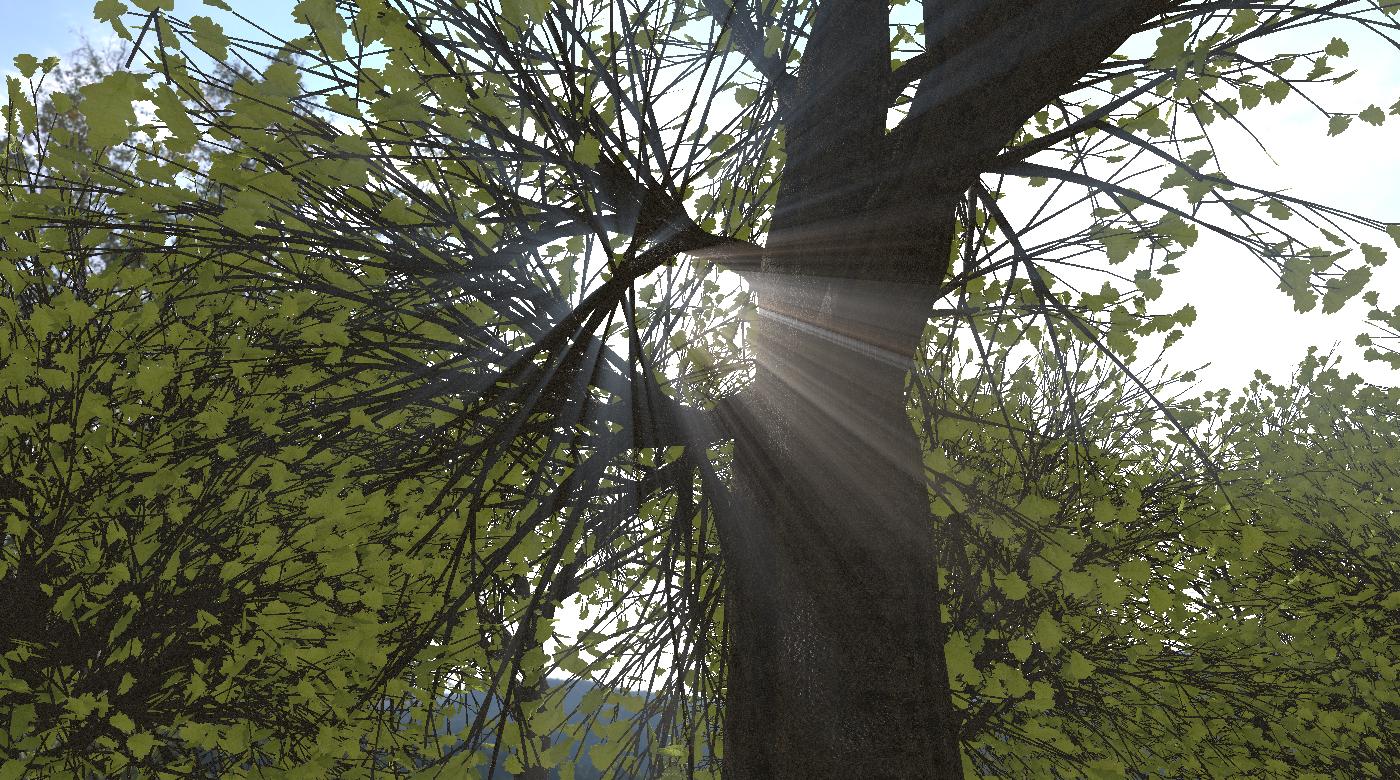}
    \end{subfigure}
    \hfill
    \begin{subfigure}[c]{0.495\linewidth}
        \centering
        \includegraphics[width=\textwidth]{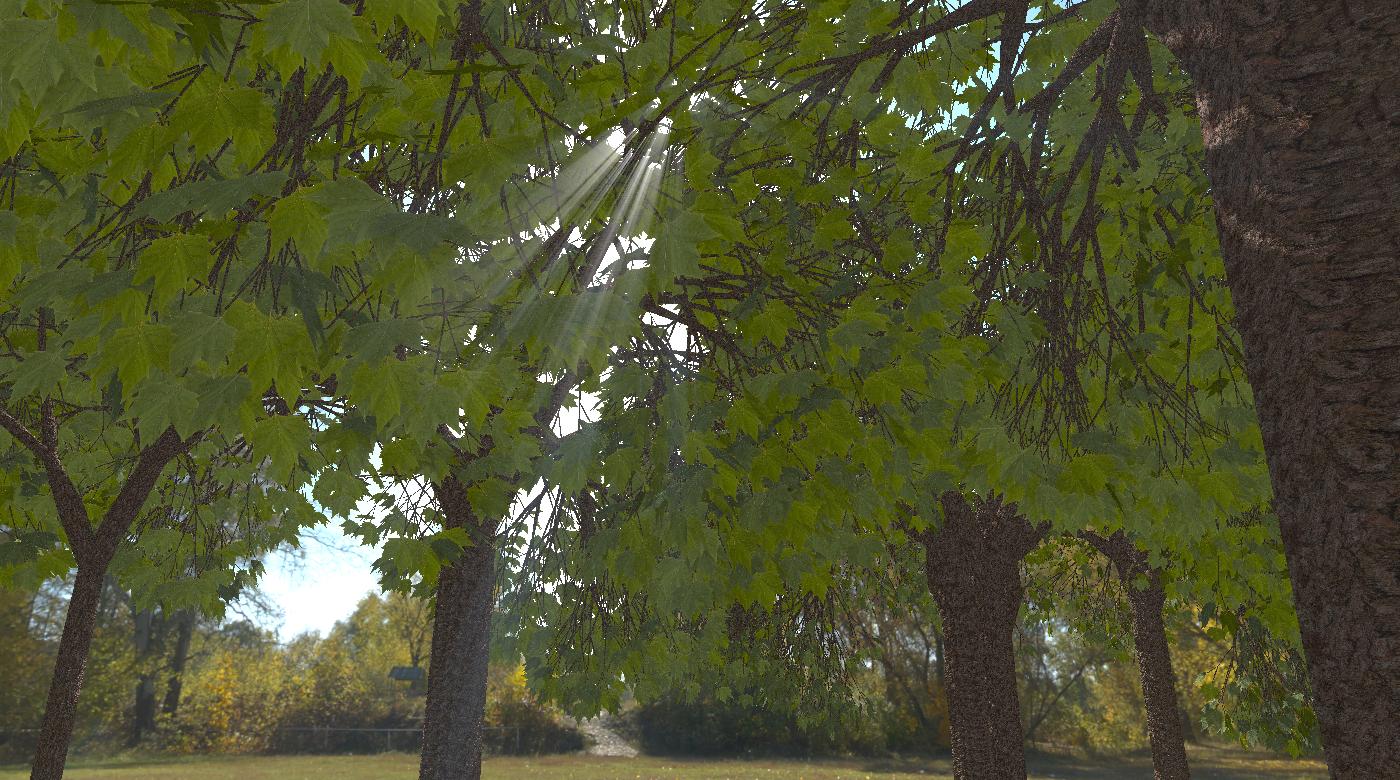}
    \end{subfigure}
    \hfill
    \begin{subfigure}[c]{0.495\linewidth}
        \centering
        \includegraphics[width=\textwidth]{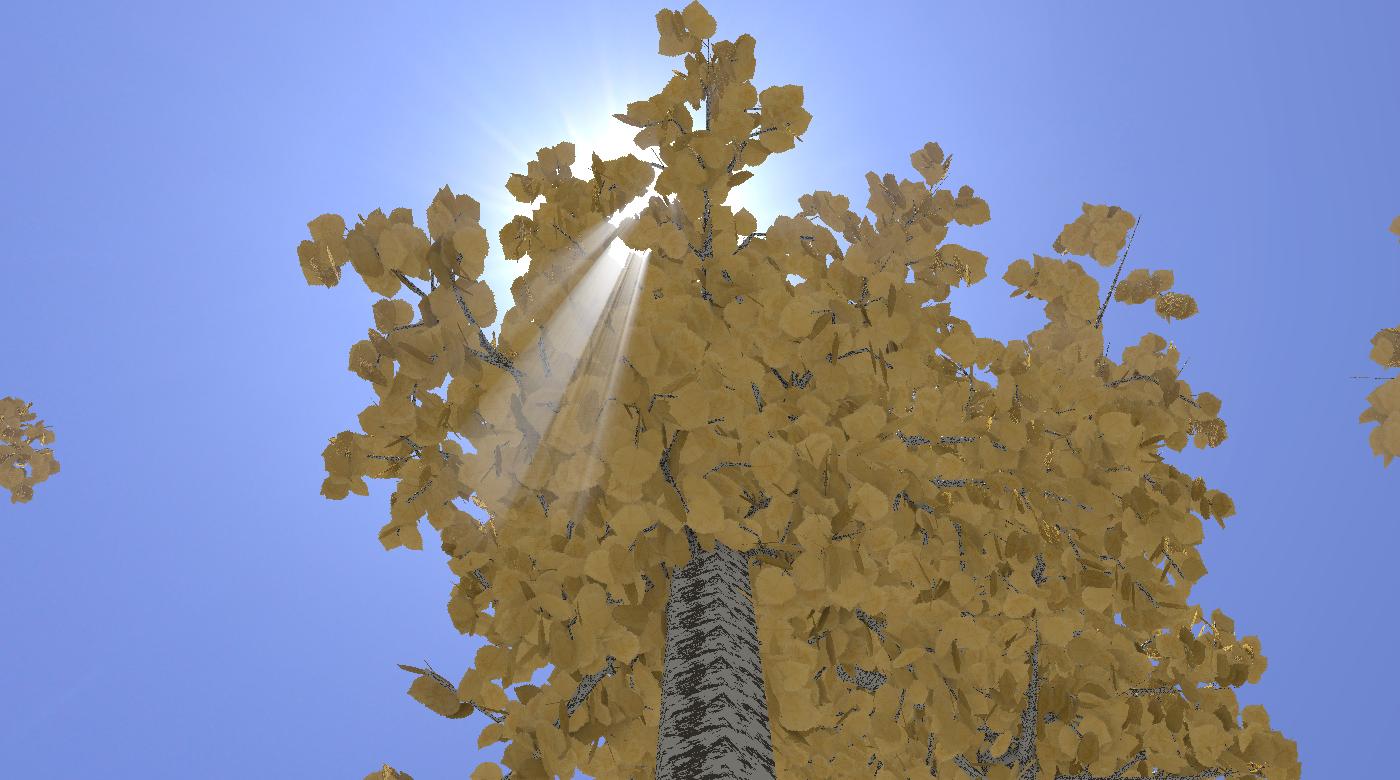}
    \end{subfigure}
    \hfill
    \begin{subfigure}[c]{0.495\linewidth}
        \centering
        \includegraphics[width=\textwidth]{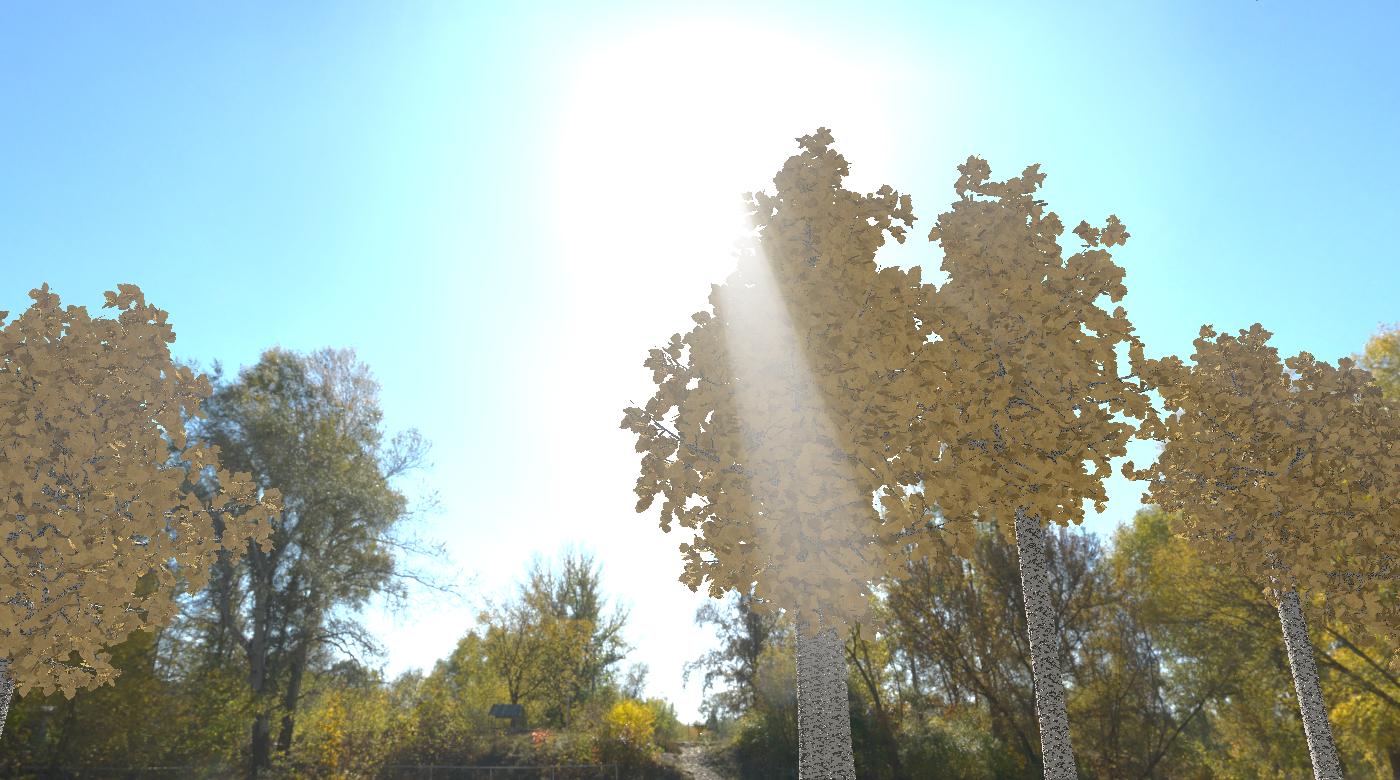}
    \end{subfigure}
    \caption{Examples of light scattering for different environment maps and positions with several tree species.}
    \label{fig:scattering}
\end{figure}

To implement volumetric light scattering (\fref{fig:scattering}),
as a screen-space post-processing shader, the system uses the method presented in \cite{Mitchell-07-GpuGems3}
adapted for use with HDRi environment maps. During the geometry pass, an occlusion map is created with the skybox fragments in full (HDR) colour, and all other fragments as black. The occlusion texture is then passed through the scattering shader which uses Equation~\ref{eq:scattering} to sample a number of points along the vector from the sun to each fragment (with an exponential decay factor). As an adaptation from the original method, the sample brightness (length of the colour vector) is clamped at a fixed value $s_{max}$ to prevent samples from appearing as bright white circles due to the high dynamic range of the skybox, giving
\begin{equation}
    L_{scatter}(\bm x) = E\sum_{i=0}^{N-1} d^i \text{min}\left(\text{occ}\left(\bm x^{ss} - \left(i+1\right) \bm \delta_s\right) ,\, s_{max}\right)
    \label{eq:scattering}
\end{equation}
where
\begin{equation}
    \bm \delta_s = \frac{\bm x^{ss} - \bm p_s^{ss}}{\rho N},
\end{equation}
$\text{occ}(\bm v)$ samples the occlusion texture at coordinate $\bm v$, $\bm x^{ss}$ is the screen-space position of $\bm x$, $\bm p_s^{ss}$ is the screen-space position of the sun, $E$ is an exposure (brightness) scalar, $N$ is the number of samples to take, $d \in [0,1)$ is the decay coefficient, and $\rho$ is the sample density (a higher density results in samples being taken from a smaller area which produces shorter, brighter rays).

This method works with any sun position or environment map and the light rays disperse well through small gaps in the leaves, with the decay parameter creating a realistic fall-off in light intensity. Furthermore, the rays update in position and appearance in a realistic manner as the camera moves around the scene. 

To enable High Dynamic Range (HDR) calculations, the system uses floating point framebuffers and then tone maps the output to return the values to the range $[0,1]$. The exposure tone mapping operator presented in \cite{Vries-20-LearnOpengl} is used to allow the user to control the tone mapping with the exposure parameter $E$
\begin{equation}
        c_{LDR} = 1 - \exp{(- E \cdot c_{HDR})}
\end{equation}
where $c_{HDR}$ is the high-dynamic range colour value.
%

The system also uses gamma correction to convert the textures from sRGB colour space \cite{sRGB} to linear colour space for calculations and then back to sRGB space for display on the screen. The gamma correction is carried out alongside tone mapping at the end of the lighting pass fragment shader.

\subsection{Rendering Optimisation} \label{Section:RenderingLOD}
The level of detail (LOD) representations for each object type were chosen so they could be pre-generated (Section~\sref{Section:Generation}). The representations for each model are stored in a hash map to enable constant-time look-ups at render time.

The LOD of a model is determined by its distance from the camera, however, to prevent calculating the distance to each one individually, a LOD is determined for each visible quadtree node (quad) based on its depth in the tree, and the models within that quad are then all represented at that LOD. Visible quads are determined recursively, where recursion stops if condition~\ref{eq:quadtreeCondition} is satisfied. This has the effect that closer quads are subdivided more, resulting in smaller (deeper) quads closer to the camera.
\begin{equation}
    \left(\lvert q_x - p_x \rvert > \tau_d w\right)\, 
     \land \,\left(\lvert q_z - p_z\rvert > \tau_d w\right)\,
    \lor\, d = D 
    \label{eq:quadtreeCondition}
\end{equation}
where $(q_x, q_z)$ is the centre of the quad on the XZ-plane, $(p_x, p_z)$ is the camera position on the XZ-plane, $w$ is the width of the quad, $d$ is the depth of the quad in the tree, $D$ is the maximum depth of the quadtree and $\tau_d$ is a threshold scalar which controls the radius of division around the camera. 

\begin{figure*}
    \captionsetup[subfigure]{labelformat=empty,skip=3pt}
    \centering 
    \begin{subfigure}[c]{0.2\linewidth}
        \centering
        \includegraphics[width=\textwidth]{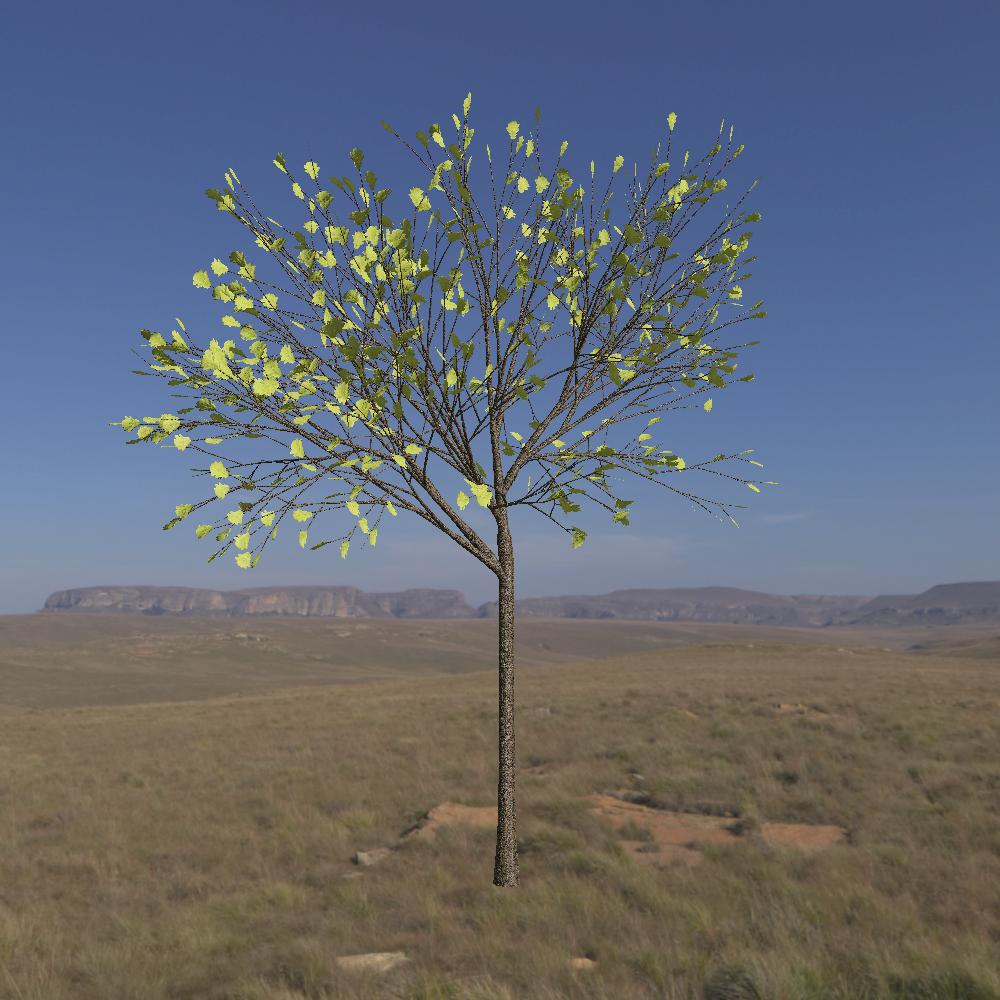}
        \caption{$n=8$}
    \end{subfigure}
    \quad
    \begin{subfigure}[c]{0.2\linewidth}
        \centering
        \includegraphics[width=\textwidth]{Figures/trees/tree-O1.jpg}
        \caption{$n=10$}
    \end{subfigure}
    \quad
    \begin{subfigure}[c]{0.2\linewidth}
        \centering
        \includegraphics[width=\textwidth]{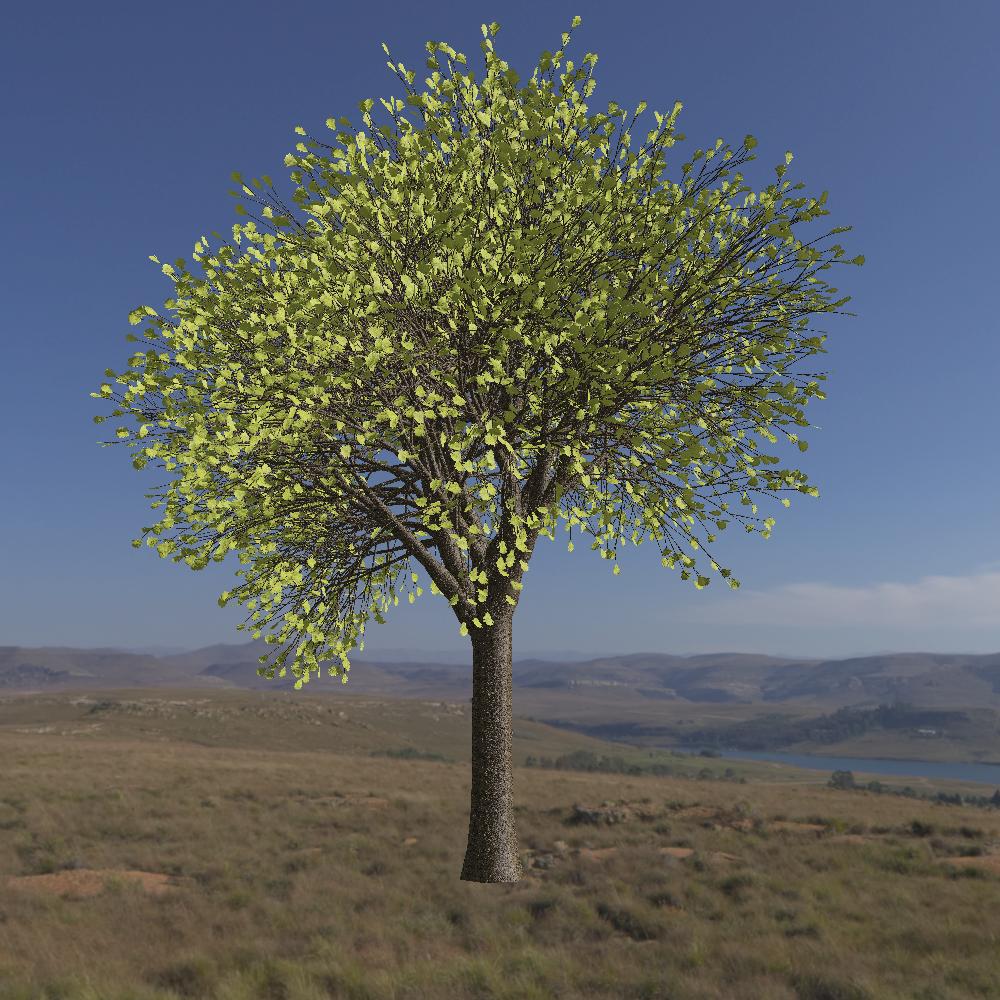}
        \caption{$n=13$}
    \end{subfigure}
    
    \caption{Three stages of growth ($n$ iterations) for the `oak' species of branching tree. The width and length of branches, as well as the number of branches and leaves, can be seen to increase as the tree `grows'.}
    \label{fig:trees-branching}
\end{figure*}

Following \cite{Gribb-01-FastExtractionViewing}, view-frustum culling is also performed at this point to check whether either of two spheres---of radius $2w$ centred at $(q_x, 0, q_z)$ and $(q_x, 2w, q_z)$---intersect with the frustum of the matrix $\bm{PV}$. Where $\bm P$ is the projection matrix for the scene and $\bm V$ is the current camera view matrix. This means that quads which are not inside the view frustum (plus an offset of 1 quad width to allow for trees which overlap quads) are not rendered.

The current LOD system has two levels (high and low)---but could be easily expanded to an arbitrary $N$ levels---with the deepest two quadtree levels ($D$ and $D-1$) rendered with high LOD, as they are closest to the camera. 

\section{Results and Discussion}


\subsection{Scene Quality} \label{Section:Quality}

The use of tree-specific L-systems\cite{Prusinkiewicz-90-algorithmicBeauty} allows for a broad range of tree species to be represented and rendered (Figure~\ref{fig:teaser}). The system generates realistic tree models with highly configurable parameters, and natural-looking leaf and branch distributions due to the varying angles and rotations. One slight issue with the use of L-systems is that branches can occasionally self-intersect in particularly dense areas. This is only noticeable at close-distance, however, and could be prevented by using an environmentally-sensitive L-System \cite{Prusinkiewicz-94-topiary}, a branch-level-parametric tree model \cite{Makowski-2019-SyntheticSilviculture}, or an intersection removal technique \cite{Xie-2016-TreeModeling}, albeit at a higher computational cost.

In the L-systems used, an increase in iterations results in increasingly later growth stages of the resulting tree (\fref{fig:trees-branching})---with the number, length and width of branches increasing---enabling tree models of differing ages to be generated from the same L-system and parameter set.

\begin{figure}
    \centering
    \begin{subfigure}[c]{0.78\linewidth}
        \centering
        \includegraphics[width=\textwidth]{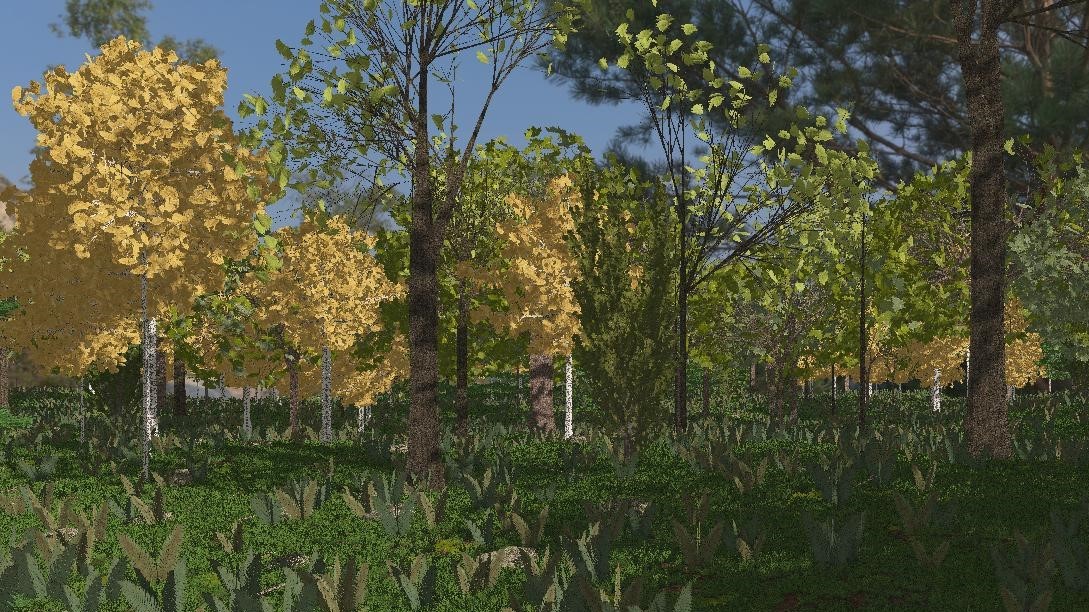}
    \end{subfigure}
    \par\smallskip
    \begin{subfigure}[c]{0.78\linewidth}
        \centering
        \includegraphics[width=\textwidth]{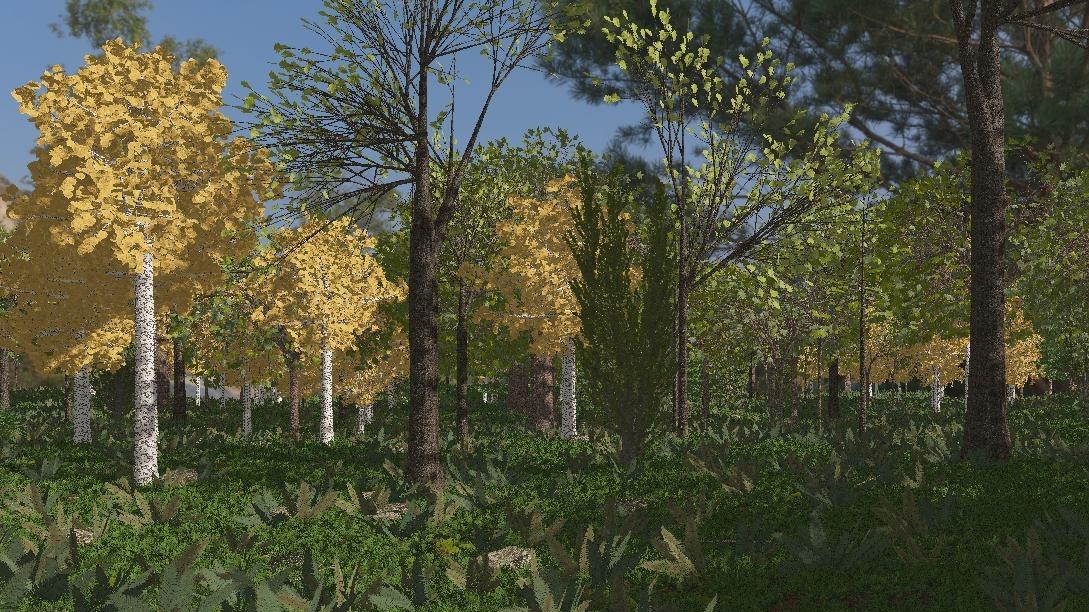}
    \end{subfigure}
    \caption{Comparison of low LOD (top) and high LOD (bottom) representations of trees and ground cover. Whilst differences between the representations are discernible at close distance, middle and far-ground objects are visually similar across both representations. }
    \label{fig:tree-lods}
\end{figure}

The LOD reduction techniques used, reduce the number of vertices in the tree models by an average of 63\% (see supplementary information for numerical breakdown)%
whilst producing visually similar results in the far and medium distances (\fref{fig:tree-lods})
with the overall tree shapes and individual branches being retained in the low LOD representation, along with the leaves exhibiting the same lighting behaviour and coverage in both representations.
\begin{figure}
    \captionsetup[subfigure]{labelformat=empty,skip=3pt}
    \centering
    \begin{subfigure}[c]{0.45\linewidth}
        \centering
        \includegraphics[width=\textwidth]{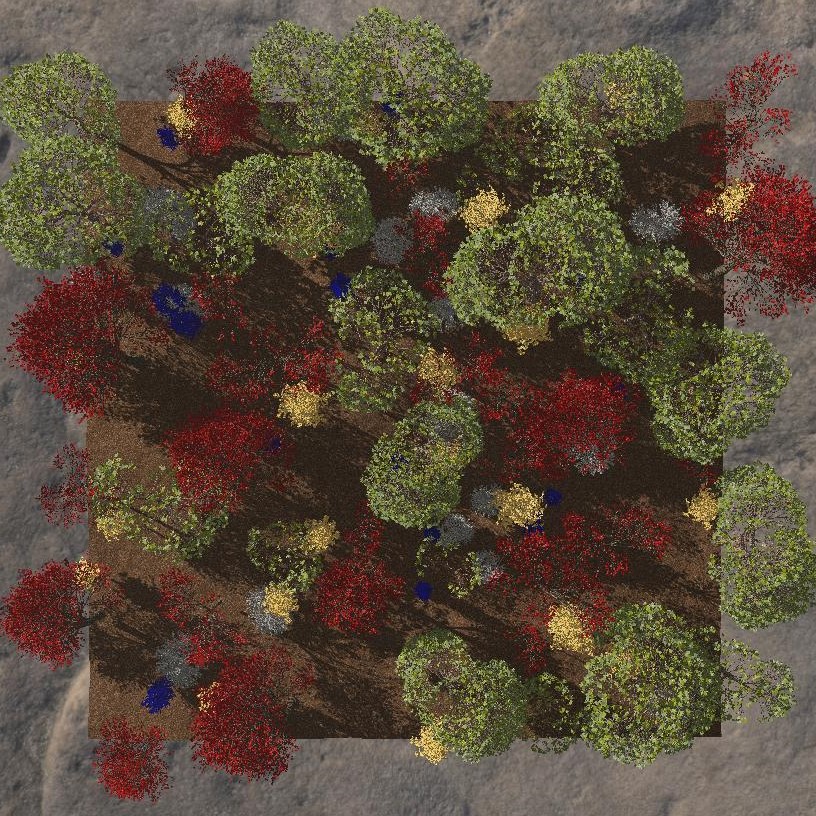}
        \caption{$n=0$}
    \end{subfigure}
    \quad
    \begin{subfigure}[c]{0.45\linewidth}
        \centering
        \includegraphics[width=\textwidth]{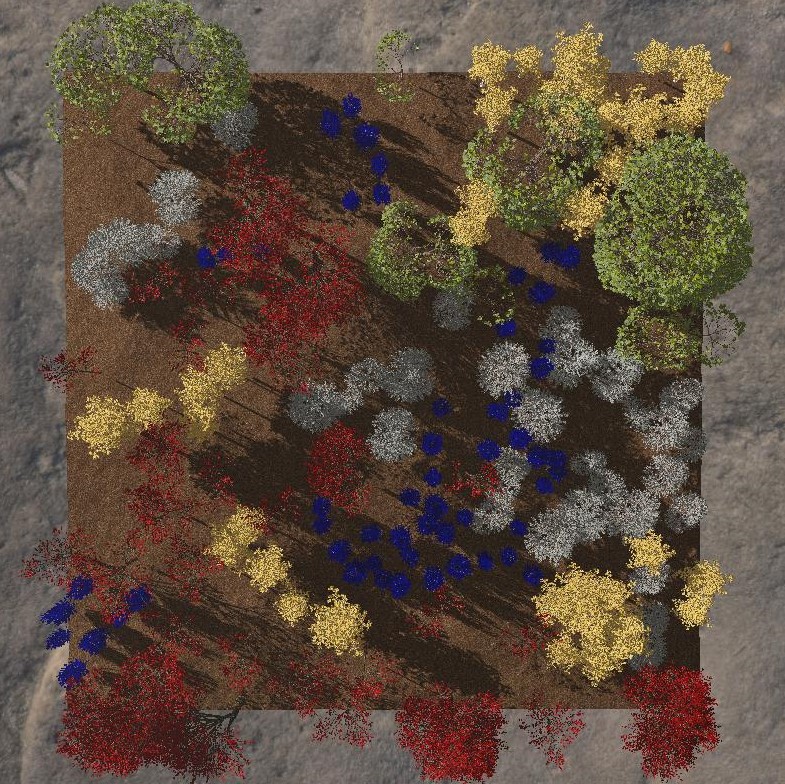}
        \caption{$n=1000$}
    \end{subfigure}
    \caption[Stages of the ecosystem distribution at different numbers of iterations]{
    The initial (random) state of the ecosystem simulation, and the distributed plants after $n=1000$ iterations.
    Clustering of plant species can be seen to emerge, and these clusters remain over time, just changing position.}
    \label{fig:dist-n}
\end{figure}%

The use of an ecosystem simulation produces tree distributions which exhibit natural features, such as plant species clustering and resource consumption, whilst also resulting in a pleasing and natural-looking forest from within. 
\begin{figure*}
    \centering
    \begin{subfigure}[c]{0.4\linewidth}
        \centering
        \includegraphics[width=\textwidth]{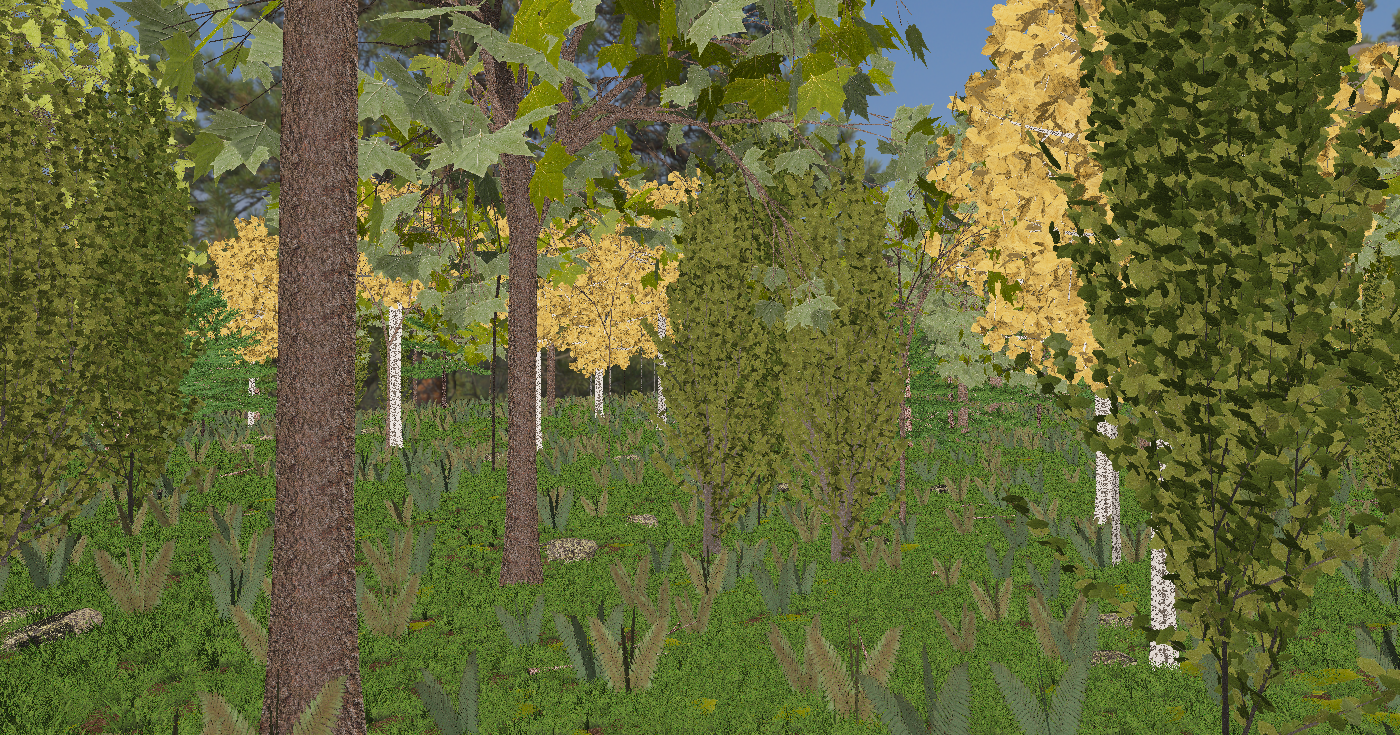}
    \end{subfigure}
    \quad
    \begin{subfigure}[c]{0.4\linewidth}
        \centering
        \includegraphics[width=\textwidth]{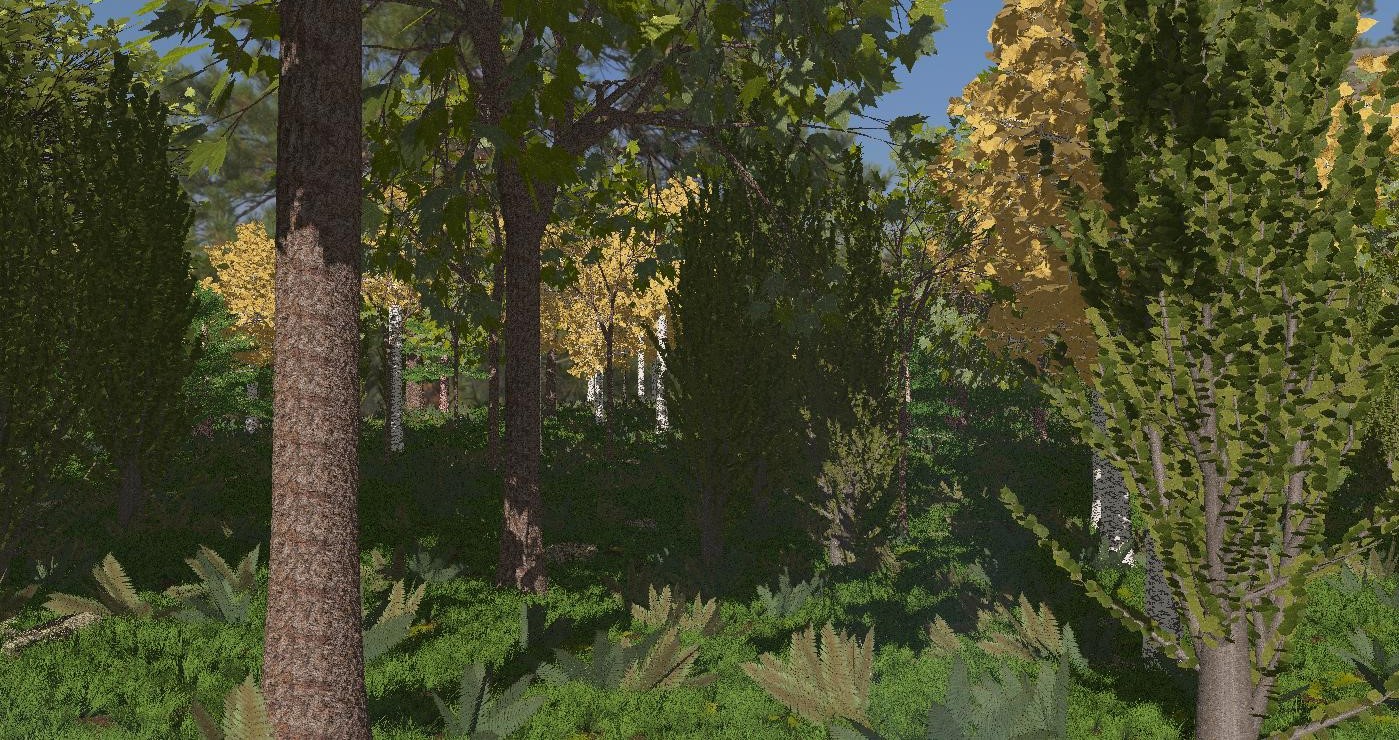}
    \end{subfigure}
    \caption{Renders of a scene with low quality (LQ) settings (left) compared to high quality (HQ) settings (right). The scene has a terrain width $w=100$, and quadtree depth $D=4$, with the LQ scene achieving an average frame render time of 62.70~ms (and maximum 66.52~ms) across three different views, compared to the HQ which required an average of 214.85~ms. }
    \label{fig:render-quality}
\end{figure*}

\begin{figure}
    \centering
    \begin{subfigure}[c]{0.64\linewidth}
        \centering
        \includegraphics[height=2.95cm]{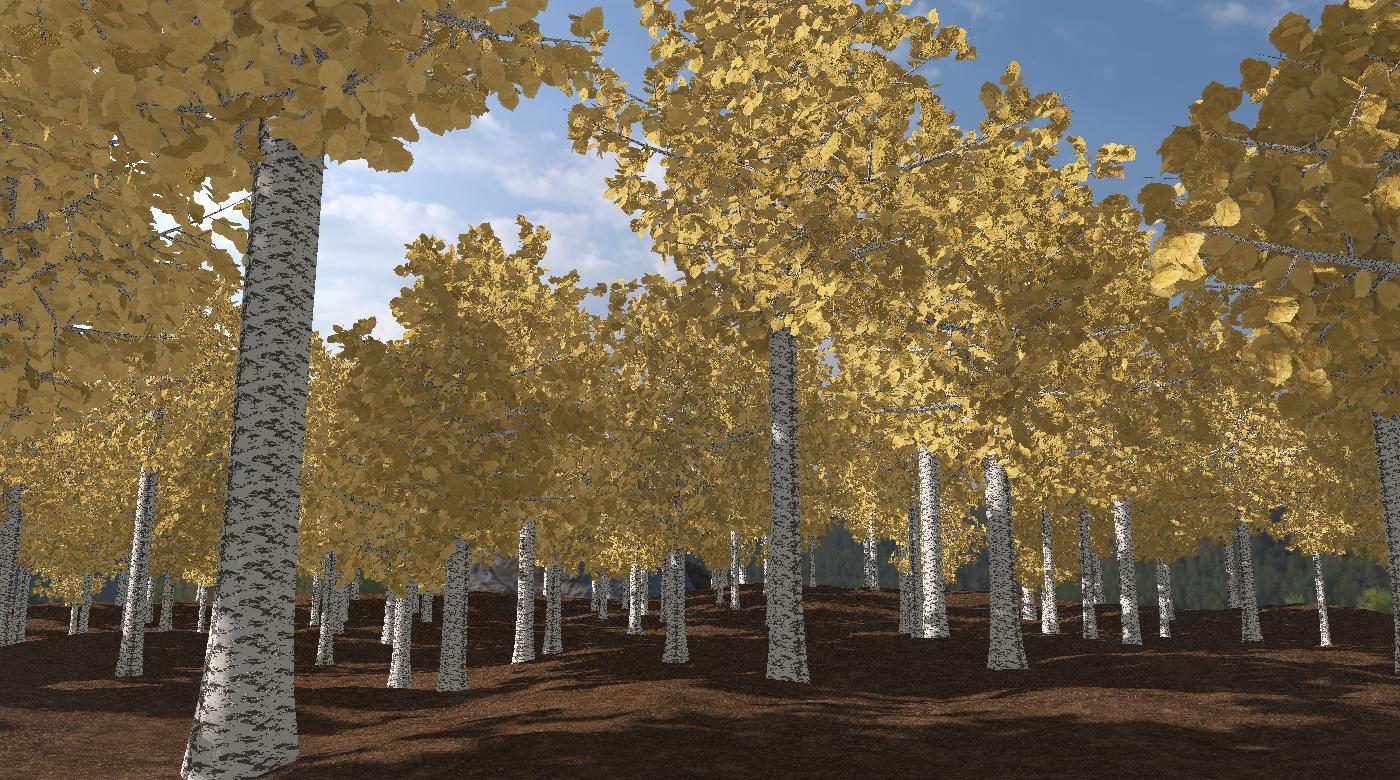}
    \end{subfigure}
    \hfill
    \begin{subfigure}[c]{0.35\linewidth}
        \centering
        \includegraphics[height=2.95cm]{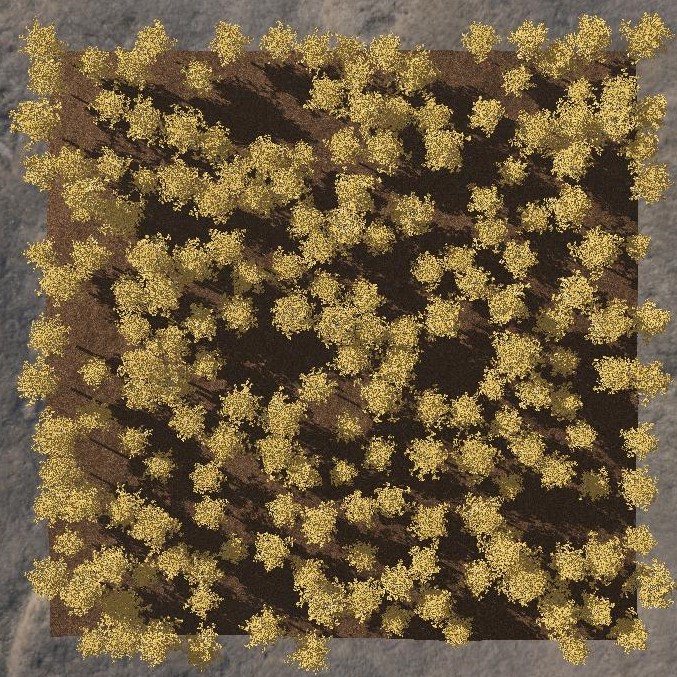}
    \end{subfigure}
    \caption[Ground view and distribution for a forest with a single species]{Ground view (left) and distribution (right) for a forest with a single species.}
    \label{fig:dist-aspen}
\end{figure}

\fref{fig:dist-n} shows the distribution produced by the simulation after 1000 iterations, where plant clustering can be seen to have emerged---unlike the initial (random) distribution in the first image.
%
%
%
Furthermore, the simulation prevents intersecting trees and larger trees inhibit the growth of other plants around them to simulate resource competition. The system also handles distributing a single plant species well (\fref{fig:dist-aspen}), resulting in a realistic distribution from both above and within the forest.

\subsection{Performance} \label{Section:Performance}
Results were measured on an Intel Core i7-8750H CPU, 2.20GHz with an NVIDIA GeForce GTX 1050 Ti Max-Q at a resolution of $1400 \times 780$~px.

The average render-time per frame across 6 runs of varying terrain sizes and quadtree depths ranged from 0.0643~s (15.5 FPS) to 0.308~s (3.24 FPS). The most time-consuming part of the render cycle is the geometry pass---taking an average of 94.8\% of the total time across the runs. This suggests that the main limiting factor for the frame rate is the number of vertices in the scene. To control this, the system parameters allow for a trade-off between quality and render time, enabling real-time interactive frame rates in the manual mode, and higher-quality output in the asynchronous input mode. Two extremes in quality are shown with their associated average frame render times in \fref{fig:render-quality}---the low-quality frames were rendered three times faster than the high-quality frames. 

\section{Conclusions}
We provide a system---released open-source\cite{NewlandsGitHub}---for generating and rendering realistic virtual forests which can be interactively explored. The system is fully configurable through the use of text files and generates the full scene before render time to reduce per-frame computation. Trees are generated with L-systems and distributed throughout the scene using an ecosystem simulation. The resulting scene is then rendered with a deferred rendering pipeline, using a Blinn-Phong shading model adapted for leaf transparency; along with high-dynamic-range tone mapping, shadow mapping and post-processing shaders for ambient occlusion and volumetric light scattering.

The system is able to generate an infinite number of tree species, terrains and overall forest scenes due to the extensive configurability of the system and flexibility of the techniques implemented. The generated tree models and distribution are naturalistic and believable, and a high-quality render can be achieved. The render performance shows that small scenes can be navigated in real-time and that, for larger scenes, the render quality can be reduced to enable real-time navigation at reasonable frame rates on a high-end laptop or PC. The system can also run asynchronously---with non-real-time output---which allows for high-quality images or video frames to be produced, or for external software (such as machine learning algorithms) to interface with the Forest Generator. For example, the tool is currently being used for training computer vision for autonomous rovers.


A future development---to increase the usability of the application---could be to adapt the system to utilise higher-level parameters \cite{Smelik-2011-SemanticConstraints, Cook-2019-GeneralAnalyticalTechniques} (which are easier to understand and control), or to apply machine learning techniques to automatically determine optimised parameter values from inputs such as photographs of real forests or user sketches \cite{Ong-05-TerrainGenerationUsing, Risi-20-IncreasingGeneralityMachine}. 

\bibliographystyle{eg-alpha-doi}
\bibliography{references}



%


\end{document}


\onecolumn
\maketitle

\section{Turtle Interpretation of L-System Modules}
Given a string of modules, the implemented turtle will ``read'' the string left to right and act on each module in the following manner:
\begin{tabularx}{\linewidth}{lX}
    $F(d)$& Move forward with a step size of $d$ and add a cross-section to the vertices list.\\
    $F(d, n_l, m_i, a_r, a_l)$& Move forward with a step size of $d$ and add a cross section to the vertices list. Distribute $n_l$ copies of sub-model with index $m_i$ around the resulting prism with radial angle $a_r$ and lift angle $a_l$.\\
    $T(v_1, v_2, v_3, e)$& Enable tropism with direction $(v_1, v_2, v_3)$ and elasticity $e$.\\
    $T(0)$& Disable tropism.\\
    $+(\delta)$& Rotate $\delta$ radians around the \texttt{up} vector.\\
    $\&(\delta)$& Rotate $\delta$ radians around the \texttt{right} vector ($\texttt{up} \times \texttt{heading}$).\\
    $/(\delta)$& Rotate $\delta$ radians around the \texttt{heading} vector.\\
    $ \$$& Reset the turtle to face in the direction $(0, 1, 0)$.\\
    $!(w)$& Set the cross section radius to $w$.\\
    $\%$& Close the current face: set the radius to 0 and insert a cross section to the vertices list.\\
    $\sim(m_i)$& Inject a copy of the sub-model with index $m_i$.\\
    $\texttt{[ }$& Push the turtle state onto the stack (for branching).\\
    $\texttt{ ]}$& Pop the turtle state off the stack and return it to the turtle. This has the effect of causing the turtle to reset its position and heading.\\
\end{tabularx}

\section{Twig L-system}

\lsysref{lst:twig} is used to generate twig models.
\begin{figure}[h!]
    \centering
\begin{lsystem}[Twig]\label{lst:twig}
    \centering
    \begin{tabular}{ LLLL }
        \rule{0pt}{4ex}    
        \omega : &\multicolumn{3}{L}{!(2)A(10, 1)} \\
        
        \rule{0pt}{4ex}    
        p_{1} :    & A(l,w)    & \xrightarrow{0.4} &
        !(w)F(l)[\&(22.5\degree)B(0.6 * l, 0.707 * w)]/(137.5\degree) \\
        &&& \qquad A(0.9 * l, 0.707 * w)\\
        
        \rule{0pt}{4ex}    
        p_{2} :    & A(l,w)    & \xrightarrow{0.6} &
        !(w)F(l)A(0.9 * l, 0.707 * w)\\
        
        \rule{0pt}{4ex}    
        p_{3} :    & B(l,w)    & \xrightarrow{0.3} &
        !(w)F(l)[+(-22.5\degree)\$A(0.9 * l, 0.707 * w)]\\
        
        \rule{0pt}{4ex}    
        p_{4} :    & B(l,w)    & \xrightarrow{0.7} &
        !(w)F(l)\\
        
    \end{tabular}
    \addcontentsline{lol}{lstlisting}{\lsysref{lst:twig} Twig}
\end{lsystem}
\end{figure}

\section{Terrain Mesh Construction Details}

To generate a ground tile of width $w$ and centre ($c_x, c_z$) with $n_v$ vertices per side, first a heightmap $h$ is constructed with $n_v + 2$ pixels on each edge. 
For each point $(i,\, j)$, 
\begin{equation}
    h[i,\,j] = \text{noise}(x',\, z') * \texttt{verticalScale}
\end{equation}
where $\text{noise}$ is the \cite{Peck-20-FastnoiseLite} implementation of 2D fractal Simplex noise and 
\begin{equation}
    \begin{gathered}
    x' = \text{lerp}\left(c_x - \left(\frac{w}{2} + \frac{w}{n_v}\right),\,\, c_x + \left(\frac{w}{2} + \frac{w}{n_v}\right),\,\, \frac{i}{n_v + 1 }\right)\,,\\
    z' = \text{lerp}\left(c_z - \left(\frac{w}{2} + \frac{w}{n_v}\right),\,\, c_z + \left(\frac{w}{2} + \frac{w}{n_v}\right),\,\, \frac{j}{n_v + 1 }\right)\,,
    \end{gathered}
\end{equation}
where $\text{lerp}$ is the standard linear interpolation function.

Once the heightmap has been generated, for each pair $(i,\, j) \in \left\{\, 1,\,\dots,\, n_v  \,\right\} \times \left\{\, 1,\,\dots,\, n_v  \,\right\}$, a vertex is constructed with position $\bm p$, normal $\widehat{\bm{n}}$, tangent $\widehat{\bm{t}}$ and texture coordinate $(x_t,\, y_t)$, resulting in a rectangular grid of faces, each rendered as 2 triangle primitives. 
%
\begin{equation}
    \bm p = \left[\texttt{round}(x),\, \texttt{round}\left(h(x,\,z)\right),\, \texttt{round}(z)\right]^T
\end{equation}
in which
\begin{equation}
    \begin{gathered}
    x = \text{lerp}\left(c_x - \frac{w}{2},\, c_x + \frac{w}{2},\, \frac{i}{n_v - 1 }\right) \\
    z = \text{lerp}\left(c_z - \frac{w}{2},\, c_z + \frac{w}{2},\, \frac{j}{n_v - 1 }\right)
    \end{gathered}
\end{equation}
$h(x,\, z) = h[i,\, j]$, and $\texttt{round}(n)$ rounds $n$ to 3d.p.\ to reduce misalignment errors between terrain tiles due to floating point errors.

A discretised, first-order approximation to the surface normal $\nabla h$ can be found using the fact that $h(x, z) = y$:
\begin{equation}
    \bm{\widehat{n}} 
    = \nabla h(x,\, z) 
    = \begin{bmatrix} 
        \dfrac{\strut\partial h}{\strut\partial x}(x, z) \\ 
        1 \\ 
        \dfrac{\strut\partial h}{\strut\partial z}(x, z)
    \end{bmatrix} 
    \approx \begin{bmatrix} 
        \dfrac{\strut h(i-1,\, j) - h(i+1,\, j)}{\strut 2 \delta x} \\ 
        1 \\
        \dfrac{\strut h(i,\, j-1) - h(i,\, j+1)}{\strut 2 \delta z}
    \end{bmatrix} 
\end{equation}
where $\delta x = \delta z = \dfrac{w}{n_v}$. Similarly
\begin{equation}
    \bm{\widehat{t}} \approx  \begin{bmatrix} 
        2 \delta x \\ 
        1 \\
        h(i+1,\, j) - h(i-1,\, j)
    \end{bmatrix} 
\end{equation}

The texture coordinates for each vertex are given by
\begin{equation}
    (x_t, \, y_t) = \left(\frac{2^{(d+s)}\, i}{n_v} ,\, \frac{2^{(d+s)}\, j}{n_v} \right)
\end{equation}
where $D$ is the maximum depth of the quadtree and $s \in \mathbb{N}$ is the value of the parameter \texttt{textureScale}. 

\section{LOD Vertex Count Reductions}
\tref{tab:tree-lods} gives the vertex counts for both the high and low LOD representations for five species of tree (sets of L-system parameters).
\begin{table}[h]
\centering
\caption[Vertex counts for high and low LOD representations of six trees]{Vertex counts for high and low LOD representations of six trees.}
\label{tab:tree-lods}
\begin{tabular}{
    cQ{0.09\textwidth}|
    Q{0.1\textwidth}Q{0.09\textwidth}|
    Q{0.1\textwidth}Q{0.09\textwidth}}
\hline
 & \multicolumn{1}{P{0.09\textwidth}|}{}  & \multicolumn{4}{P{0.4\textwidth}}{Number of Vertices} \\ \cline{3-6}
\multicolumn{1}{c}{Species} & \multicolumn{1}{P{0.09\textwidth}|}{Number of Edges} & \multicolumn{1}{P{0.09\textwidth}}{High LOD Branches} & \multicolumn{1}{P{0.09\textwidth}|}{High LOD Canopy} & \multicolumn{1}{P{0.09\textwidth}}{Low LOD Branches} &
\multicolumn{1}{P{0.09\textwidth}}{Low LOD Canopy} 
\\ \hline
\multicolumn{1}{c}{Aspen} & 6 & 115992 & 47708 & 38664 
& 23852
\\ 
\multicolumn{1}{c}{Branching}  & 6 & 787128 & 41304 & 262376
& 20652 
\\ 
\multicolumn{1}{c}{Oak}  & 6 & 611448 & 28580 & 203816
& 14288 
\\ 
\multicolumn{1}{c}{Pine}  & 8 & 52288 & 177296 & 13072 
& 88648 
\\ 
\multicolumn{1}{c}{Poplar}  & 8 & 136096 & 235880 & 34024 
& 117940 
\\ \hline
\multicolumn{1}{c}{} & Avg. & 340590 & 106153 & 110390 & 53076 \\ \hline
\end{tabular}
\end{table}

\bibliographystyle{eg-alpha-doi}
\bibliography{references}
